\documentclass[preprint2]{aastex61}

\usepackage{graphicx}
\usepackage{epstopdf}
\usepackage{lipsum}
\usepackage{amsmath}
\usepackage{amssymb}
\usepackage{float}
\usepackage{booktabs}
\usepackage{rotating}

\received{}
\accepted{}
\published{}

\submitjournal{ApJ}

\shorttitle{Revised Distances to SNRs}
\shortauthors{Ranasinghe \& Leahy}

\begin{document}

\title{Revised Distances to 21 Supernova Remnants}

\author{S. Ranasinghe and D.A. Leahy }
\affil{Department of Physics $\&$ Astronomy, University of Calgary, Calgary,
Alberta T2N 1N4, Canada}

%% Note that the \and command from previous versions of AASTeX is now
%% depreciated in this version as it is no longer necessary. AASTeX 
%% automatically takes care of all commas and "and"s between authors names.

%% AASTeX 6.1 has the new \collaboration and \nocollaboration commands to
%% provide the collaboration status of a group of authors. These commands 
%% can be used either before or after the list of corresponding authors. The
%% argument for \collaboration is the collaboration identifier. Authors are
%% encouraged to surround collaboration identifiers with ()s. The 
%% \nocollaboration command takes no argument and exists to indicate that
%% the nearby authors are not part of surrounding collaborations.

%% Mark off the abstract in the ``abstract'' environment. 
\begin{abstract}
We carry out a comprehensive study of HI 21 cm line observations and $^{13}$CO  line observations
of 21 supernova remnants (SNRs). The aim of the study is to search for HI absorption features 
to obtain kinematic distances in a consistent manner. 
The 21 SNRs are in the region of sky covered by the Very Large Array Galactic Plane Survey (HI 21 cm  observations)
and  Galactic Ring Survey ($^{13}$CO  line observations).
We obtain revised distances for 10 SNRs based on new evidence in the HI and $^{13}$CO observations.
We revise distances for the other 11 SNRs based on an updated rotation curve and new error analysis. 
The mean change in distance for the 21 SNRs is $\simeq25\%$, i.e. change of 1.5 kpc compared to a mean distance for
the sample of 6.4 kpc. This has a significant impact on interpretation of the physical state of these SNRs. For example, using a 
Sedov model, age and explosion energy scale as the square of distance, and inferred ISM density scales as distance.

\end{abstract}

\keywords{ISM: supernova remnants - radio continuum: ISM - radio lines: ISM}

\section{Introduction} \label{sec:intro}

\indent 
Supernova remnants (SNRs) play an important role in determining the state of the interstellar medium of the Galaxy,
as described in, e.g., the reviews by \cite{2005Cox} and \cite{2001Ferriere}.
However, in order to understand SNRs, reliable %accurate 
distances are necessary. 
Other objects, including HII regions and molecular clouds, are often associated with SNRs. 
To determine whether any given association is real, the SNR distance is needed. \\
\indent  One common method to find the distance to SNRs is to obtain kinematic distance from analysis of HI absorption spectra. 
This method has yielded reliable distance estimations, but it has been mainly applied to brighter SNRs. 
The HI absorption spectra of fainter SNRs are usually noisy. This is caused by real fluctuations, i.e. HI emission occurring
at many random positions and velocities in both source and background regions used to create the absorption spectrum.
For bright SNRs, these fluctuations are small compared to the absorption signal but for faint SNRs they are comparable. 
For the fluctuations in HI emission, the HI and continuum images are poorly correlated, whereas for real absorption there
is a clear match between the absorption signal (decreased HI intensity) and the continuum emission. 
Thus careful investigation of the HI channel images is essential to assess the reality of any features which are seen in the HI absorption spectrum.
Here we follow the analysis methods described by \cite{Leahy2010} and  \cite{2017Ranasinghe}. \\
\indent 
The sample of SNRs that we consider are those that cover the region of the VLA (Very Large Array) Galactic Plane Survey (VGPS) \citep{Stil}.
There are 59 SNRs in the sky area covered by the VGPS survey.
Of these, we have previously studied 10 SNRs without previous distance determinations, including SNRs G31.9+0.0 and G54.4‑0.3
\citep{2017Ranasinghe},  4 SNRs with new molecular cloud associations \citep{2017RanasingheLeahy}, 
and 4 SNRs without molecular cloud associations \citep{2017RLT}.
In this work we analyze data for SNRs with previously published distances and for which the HI line data quality is high enough for analysis.
This results in 21 SNRs in our sample presented here. 
For 10 of these SNRs, we revise previous distances based on new evidence from the HI and  $^{13}$CO data.
For the remaining 11 SNRs, we confirm published kinematic velocities and revise distances using an updated error analysis and a more recent Galactic rotation curve. 
In Section \ref{sec:DA}, we present a brief description of the data, construction of HI absorption spectra and kinematic distances. 
The details for each of the 10 SNRs with new evidence are given in Section \ref{sec:results}. 
The discussion and the final table summarizing our work for all 21 SNRs is described in Section \ref{sec:dissummary}.

\section{Data and Analysis} \label{sec:DA}

The  HI line data and $^{13}$CO line data are from the VGPS Survey \citep{Stil} 
and the Galactic Ring Survey of the Five College Radio Astronomical Observatory (FCRAO) \citep{Jackson}.
The method for making spectra and following analysis is from \cite{Leahy2010} and \cite{2017Ranasinghe}.\\
\indent For kinematic distances, a reliable rotation curve is necessary. 
We adopt the universal rotation curve (URC) of \cite{Persic} with the \cite{Reid2014} parameters: 
Galactocentric radius of $R_{0} = 8.34 \pm 0.16$ kpc and orbital velocity of the sun $V_{0} = 241 \pm 8$ km s$ ^{-1}$. 
We verified the tangent point velocities predicted by this rotation curve as follows. 
For each SNR we obtained the HI emission spectrum adjacent to the SNR.
We fit the high velocity part of the HI emission spectrum by a model spectrum of Galactic HI emission. 
The model spectrum was constructed by integrating, along the line of sight, the HI density as a function of position and velocity. The velocity of HI includes circular rotation 
and local Gaussian velocity dispersion. 
The adjustable parameters of interest from the fit were the circular velocity at the tangent point (thus tangent point velocity) and the velocity dispersion.
The tangent point velocity obtained from the fit to the data agreed with that from the URC model within 3 km s$ ^{-1}$, except for 2 SNRs.
For those 2 cases, the model fit and the URC tangent point velocities differ from significantly (see sections  \ref{G49} and  \ref{G54}).
The tangent point velocities from the fit to the data should be more reliable than from the URC, because the URC assumes an
axisymmetric Galaxy, whereas the fit to the data does not.  
Thus the data fit allows for non-axisymmetric effects such as velocity perturbations due to spiral arms or the Galactic bar. 
We chose to use the tangent point velocities from the fits to the HI emission spectrum for those two cases. 

To determine the tangent point error we use the error in V$_{r}$ = 5.3 km s$ ^{-1}$ \citep{2017RanasingheLeahy}. 
For the errors in the distances other than the tangent point, we follow the method presented by \cite{2017RanasingheLeahy}. 

\section{Results} \label{sec:results}

For 10 of the 21 SNRs we obtain new evidence leading to different distances than previously published values. 
These SNRs are:  G$20.0 -0.2$, G$23.6 +0.3$, G$27.4 +0.0$, G$33.6 +0.1$, G$34.7 -0.4$,  G$39.2 -0.3$, 
G$43.3-0.2$, G$46.8-0.3$, G$49.2-0.7$ and  G$54.1+0.3$.    
A brief overview of previous work and the arguments and relevant data for the new distances for each of these 10 SNRs are presented in the following sections.
For the 11 SNRs for which we verify velocities from previous work, our results are presented only in the final table of distances.
The distances for these 11 SNRs are different than previously published distances because we use an updated rotation curve. 

\subsection{G$20.0 -0.2$} \label{G20}

\indent G$20.0 -0.2$ was classified as a Crab-like supernova remnant \citep{BeckerHelfand1985}. 
This was based on the flat radio spectral index, filled-center morphology and the presence of substantial polarization at 6 cm. 
To the north-left of the SNR, lies a compact and bright source, identified as the ultra-compact HII region GAL $20.08 -0.14$ \citep{Wood1989}. 
\cite{Anderson20091} declared there is no ambiguity in position or velocity of this HII region. The velocity of 42.5 km s$^{-1}$ yields the far side distance of 12.6 kpc (\cite{Petriella2013}, \cite{Anderson22009}).\\
\indent \cite{Petriella2013} presented the first X-ray study, using archival Chandra observations to establish whether G$20.0 -0.2$ is purely plerionic or composite. 
They discovered diffuse X-ray emission with a non-thermal spectrum toward the center of the radio emission, so it is more likely the former.
Using the $^{13}$CO (J = 1-0) data of \cite{Jackson}, \cite{Petriella2013} presented evidence of a molecular cloud extending from 62 to 72 km s$^{-1}$  
associated with the SNR. 
Using the velocity of 66 km s$^{-1}$  the cloud is quoted to be at 4.5 or 11.5 kpc, and their analysis of the HI absorption favored the near distance.\\

\begin{figure}[ht!]
\centering
\includegraphics[width=\columnwidth]{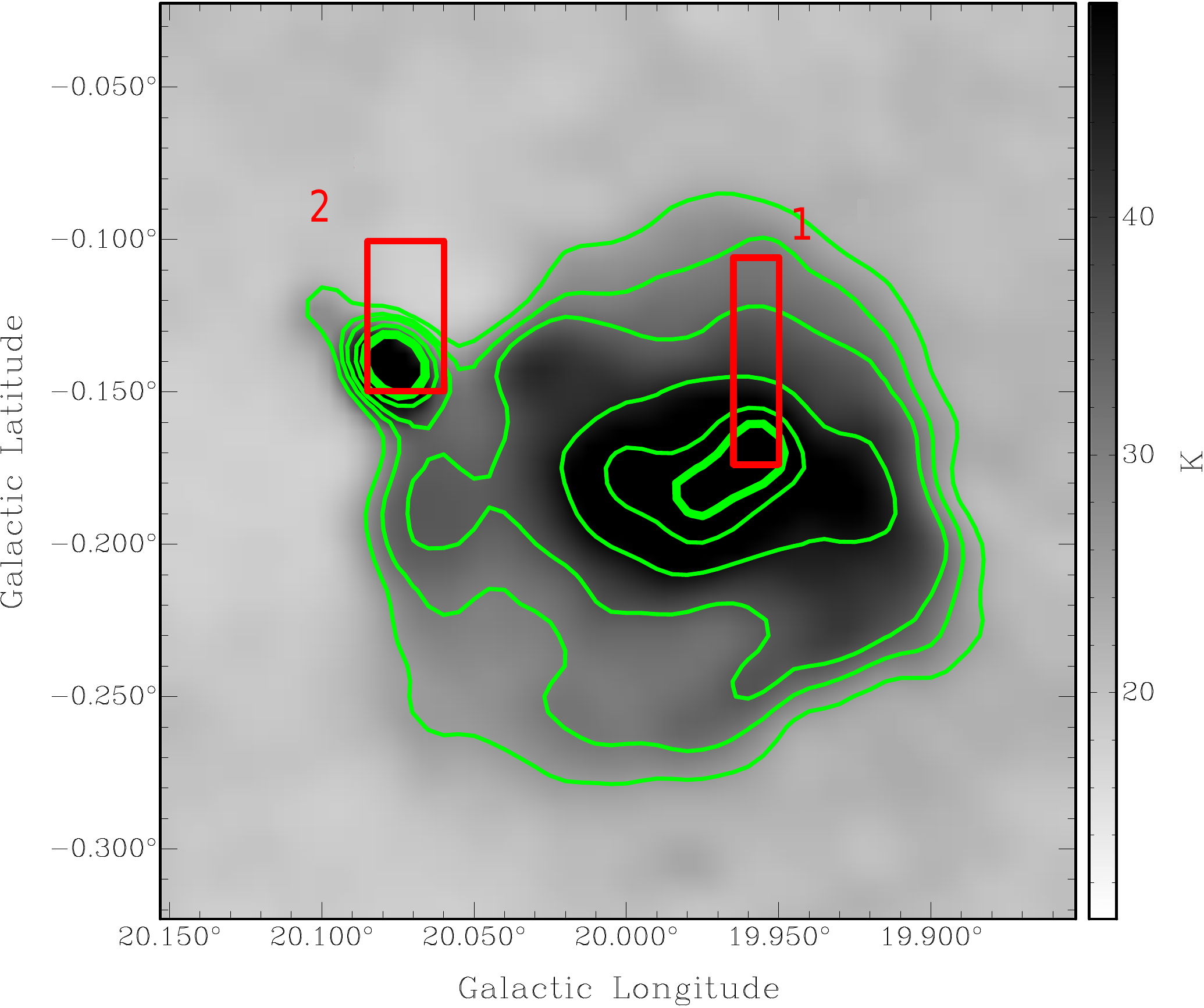}
\caption {SNR G$20.0 -0.2$ 1420 MHz continuum image. Contour levels (green): 25, 30, 35, 45, 55 and 60 K.  The red boxes are the regions used to extract HI and $^{13}$CO source and background spectra.}
\label{fig:1}
\end{figure}

\begin{figure*}[ht!]
\includegraphics[scale=0.35]{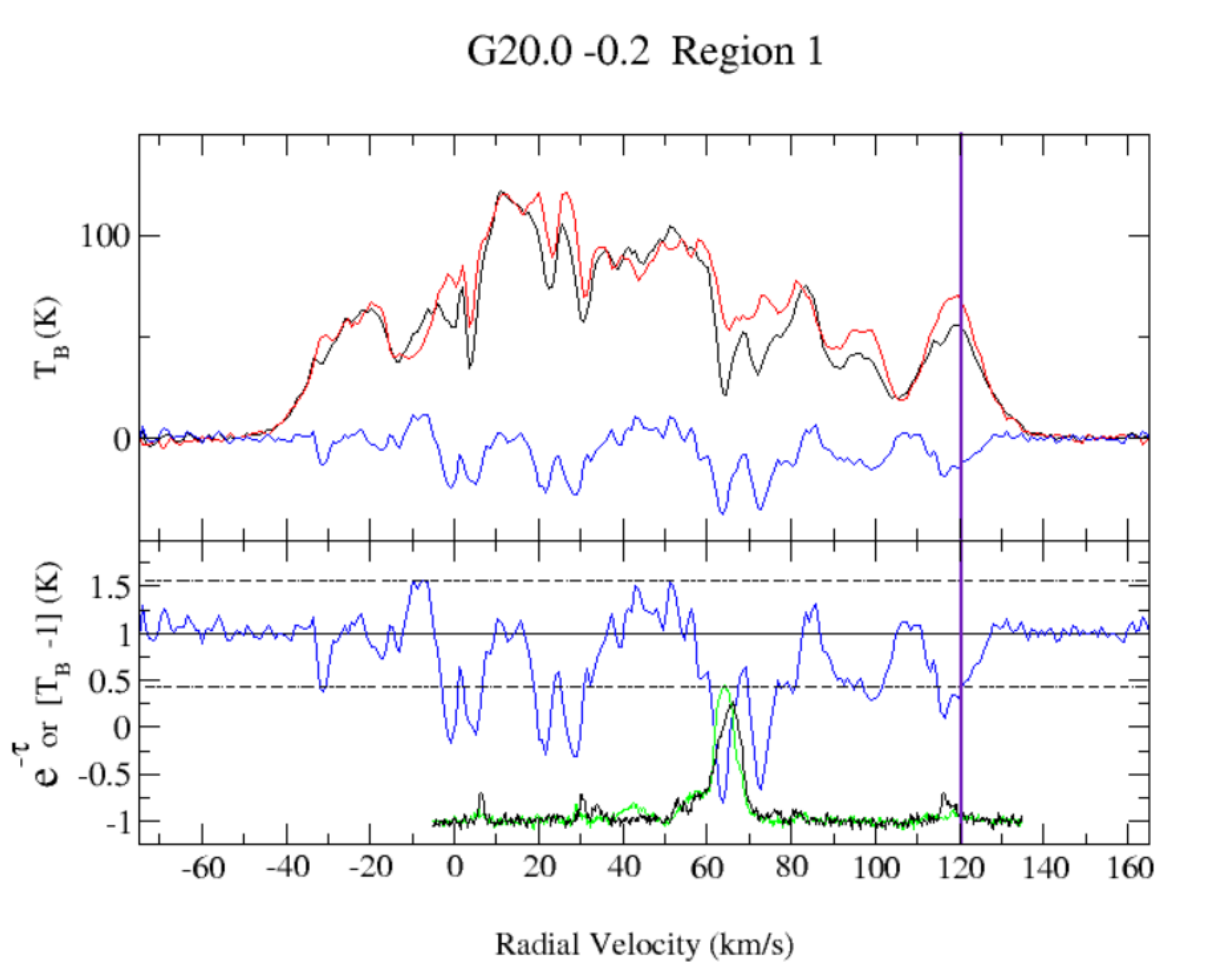}
\includegraphics[scale=0.35]{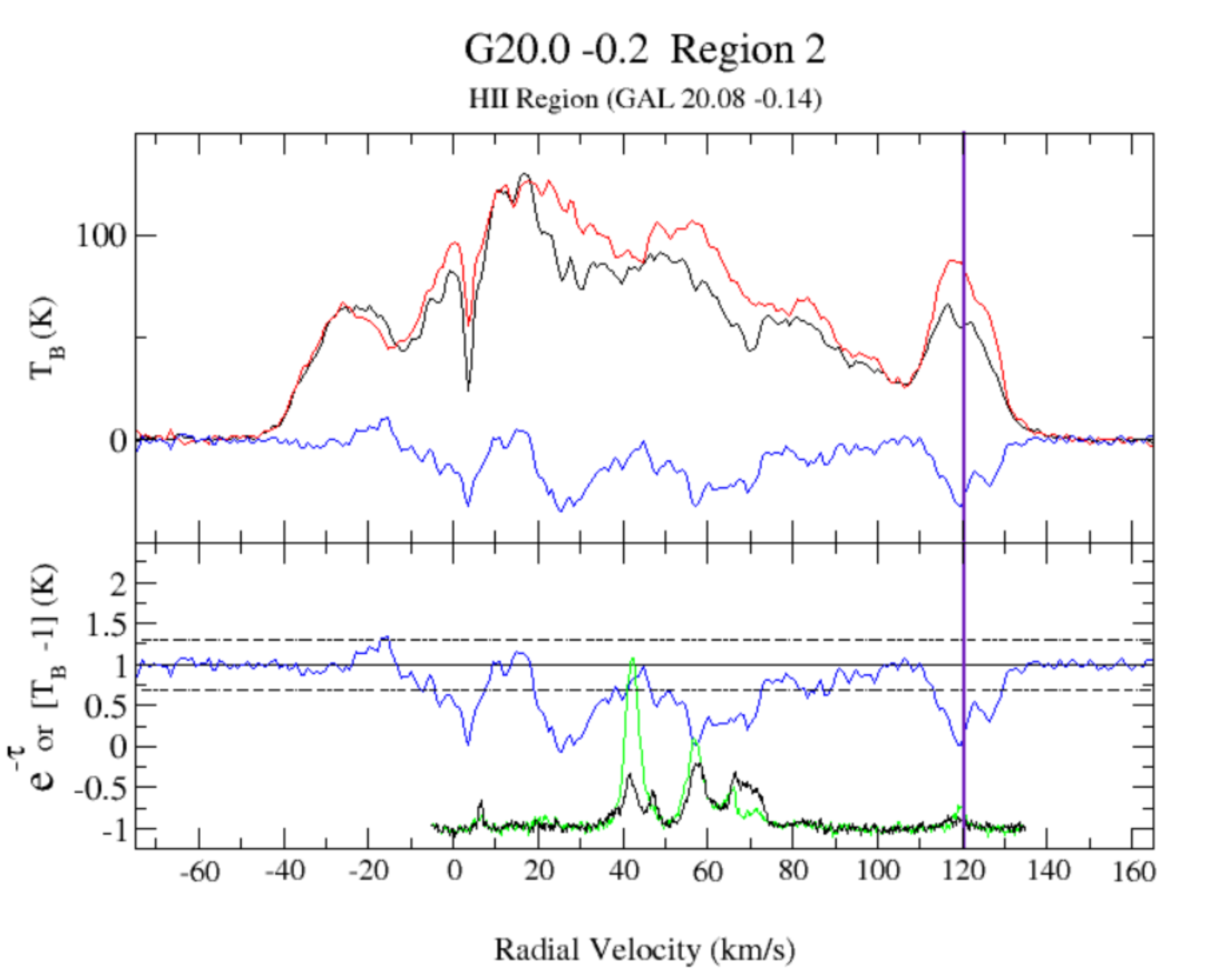}
%\plottwo{Figure2A.png}{Figure2B.png}
\caption{SNR G$20.0 -0.2$ (left) and HII region GAL $20.08 -0.14$ (right) spectra. Upper half of panel: HI emission spectrum (source: black, background: red and difference: blue). Lower half of panel: HI absorption spectrum (e$^{-\tau}$, blue), $^{13}$CO source (green) \& background (black) spectra (T$_{B}$, offset by subtracting 1 K), $\pm2 \sigma$ noise level of the HI absorption spectrum (dashed line) and tangent point velocity (purple vertical line).}
\label{fig:2}
\end{figure*}

\begin{figure*}[ht!]
\includegraphics[width=\textwidth]{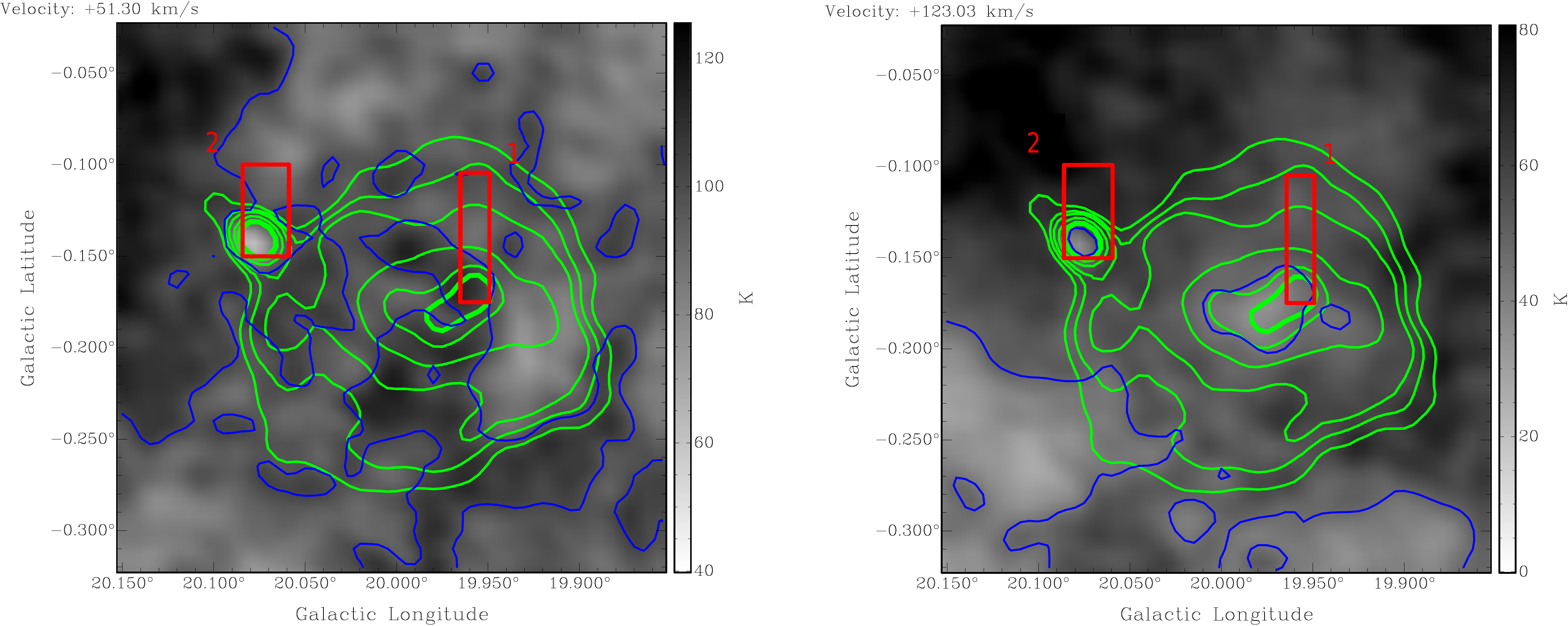}
\caption{G$20.0 -0.2$ HI channel maps +51.30 and +123.03  km s$^{-1} $. HI contour levels (blue): 46 and 100 K respectively. Continuum contour levels (green): 25, 30, 35, 45, 55 and 60 K.}
\label{fig:3}
\end{figure*}

\indent Figure \ref{fig:1} shows the continuum image with region 1, chosen for HI absorption spectrum extraction. 
The HII region GAL $20.08 -0.14$ spectrum (Region 2) was used as a comparison. 
The HI spectra of the SNR and the HII region both show absorption present up to the tangent point (Figure \ref{fig:2}). 
The spectrum of the SNR has a number of false features. 
These can be identified as caused by HI clouds in the source or background regions seen in the channel maps.
For example, the 51.30 km s$^{-1}$ image (Figure \ref{fig:3} left panel) shows a cloud in the source region (box 1) which results in $e^{-\tau} > 1$.
Conversely a false absorption feature could be seen if the higher intensity HI lies in the background region. 

The absorption up to the tangent point is confirmed by the HI channel maps (Figure \ref{fig:3} right panel).
Studying both spectra and the HI channels of the SNR, it is seen that there is clear absorption from $\sim$110 km s$^{-1}$  to the tangent point velocity of $\sim$123 km s$^{-1}$. 
The absorption features in the channel maps correlate morphologically to the brightest regions of the SNR and are highly likely to be real. 
 Thus the lower limit distance to the SNR is the tangent point distance of 7.8 kpc. 

From the $^{13}$CO channel maps, the molecular clouds  associated with the SNR are evident between 61.10 and 71.51 km s$^{-1}$ (see \cite{Petriella2013} Figure 5). 
Adopting 66.40 km s$^{-1}$ as the central velocity of the molecular cloud associated with the SNR, the distance to G$20.0 – 0.4$ is estimated to be 11.2 kpc.\\ 
\indent The molecular cloud associated with the HII region GAL $20.08 -0.14$, is consistent with Anderson et al. (2009) analysis and has velocity of 42.5 km s$^{-1}$,
  yielding a distance of 12.4 kpc. The HII region is not associated with the SNR.

\subsection{G$23.6 +0.3$}  \label{G23}
        
\begin{figure}[ht!]
\centering
\includegraphics[width=\columnwidth]{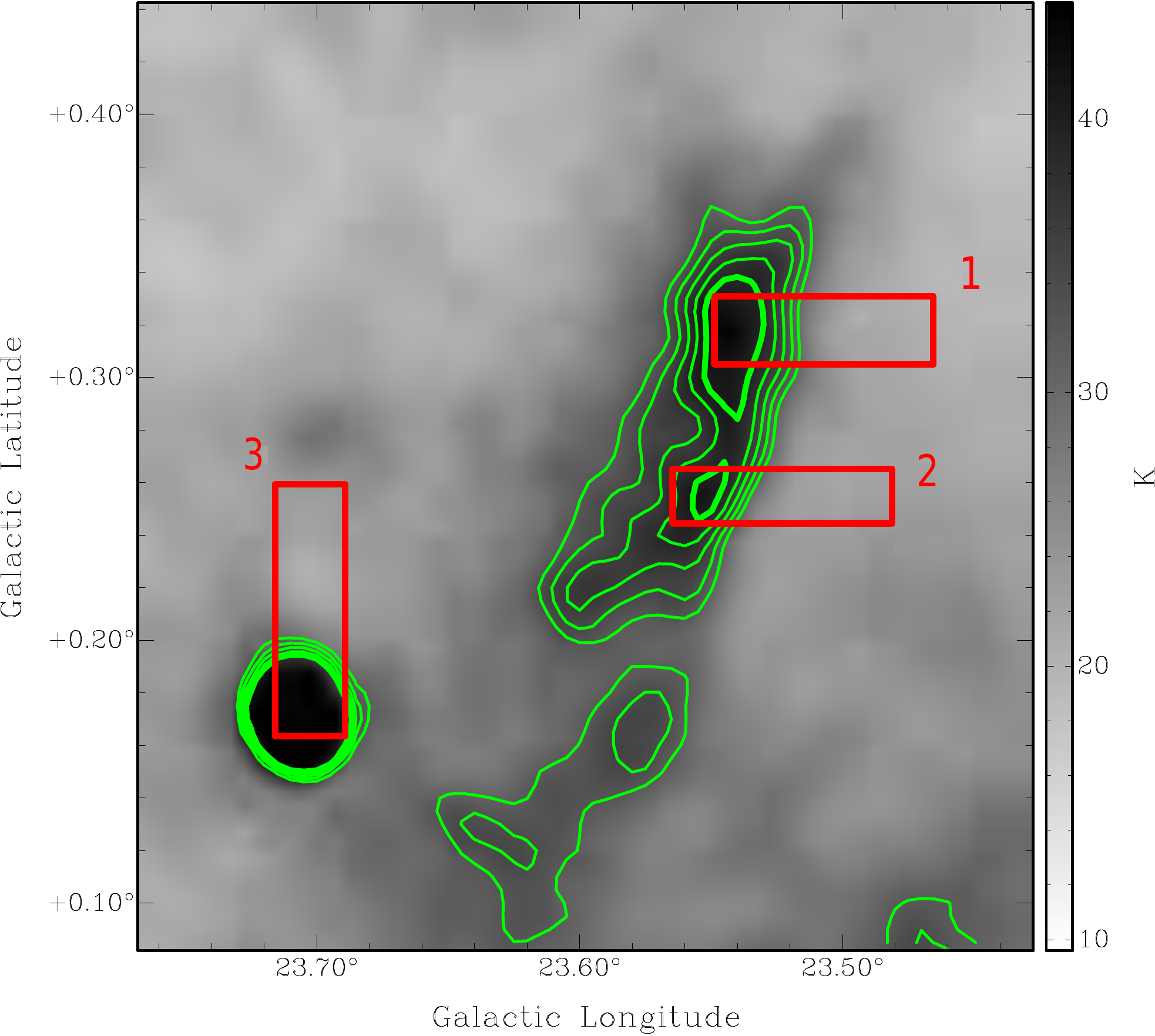}
\caption {SNR G$23.6 +0.3$ 1420 MHz continuum image. Contour levels (green): 32, 34, 36, 38 and 40 K. The red boxes are the regions used to extract HI and $^{13}$CO source and background spectra.}
\label{fig:4}
\end{figure}

\begin{figure*}[ht!]
  \includegraphics[scale=0.35]{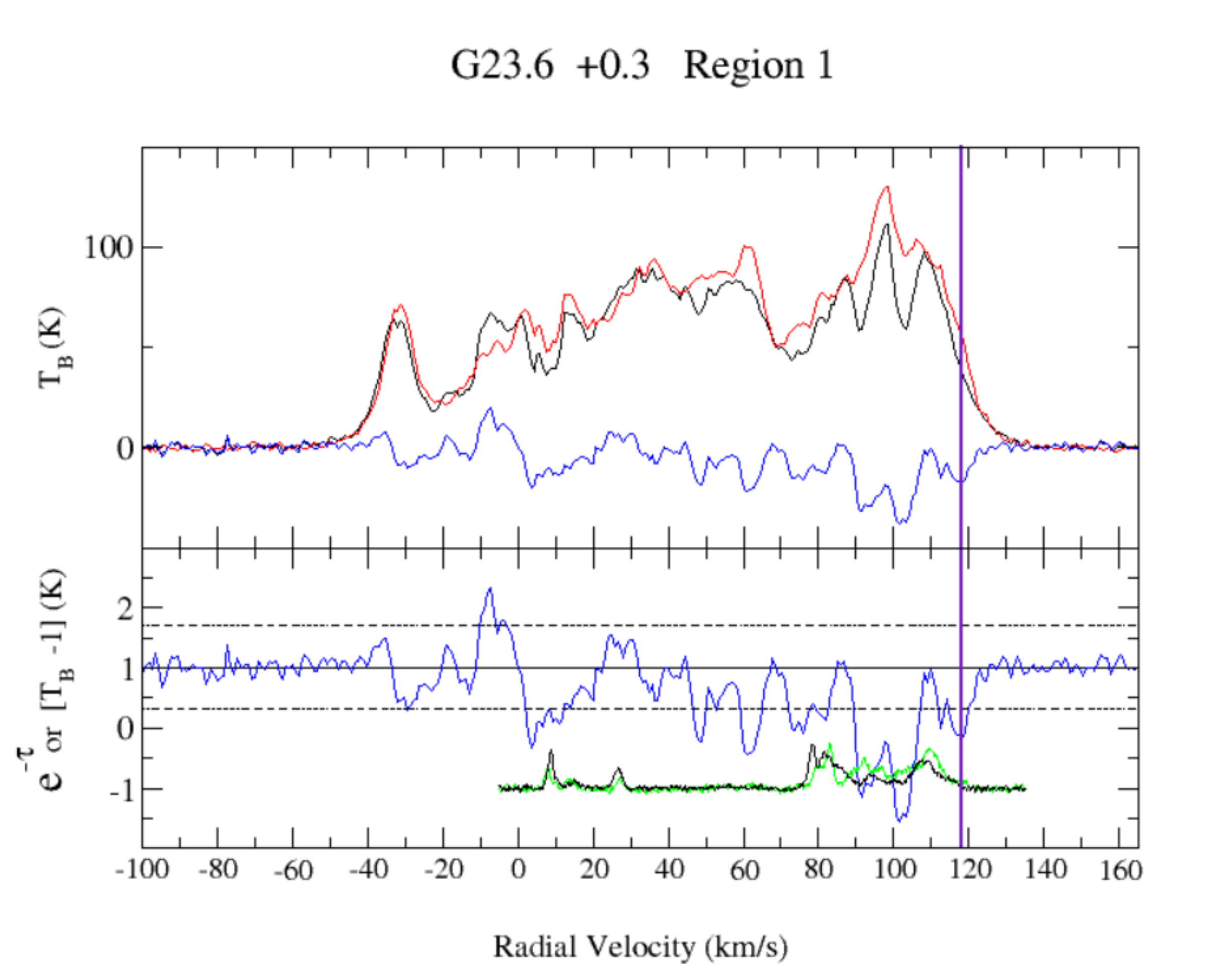}
  \includegraphics[scale=0.35]{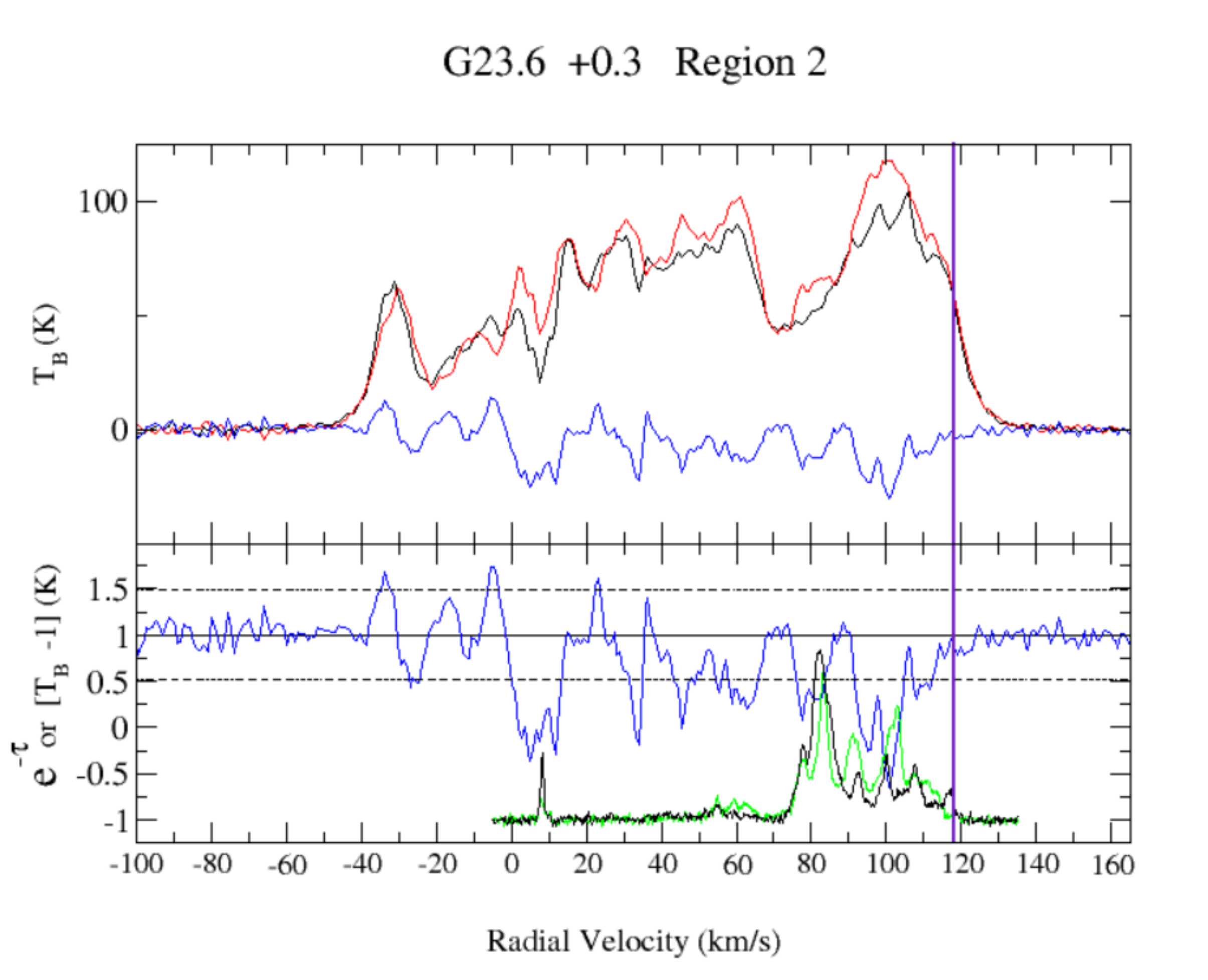}
  \includegraphics[scale=0.35]{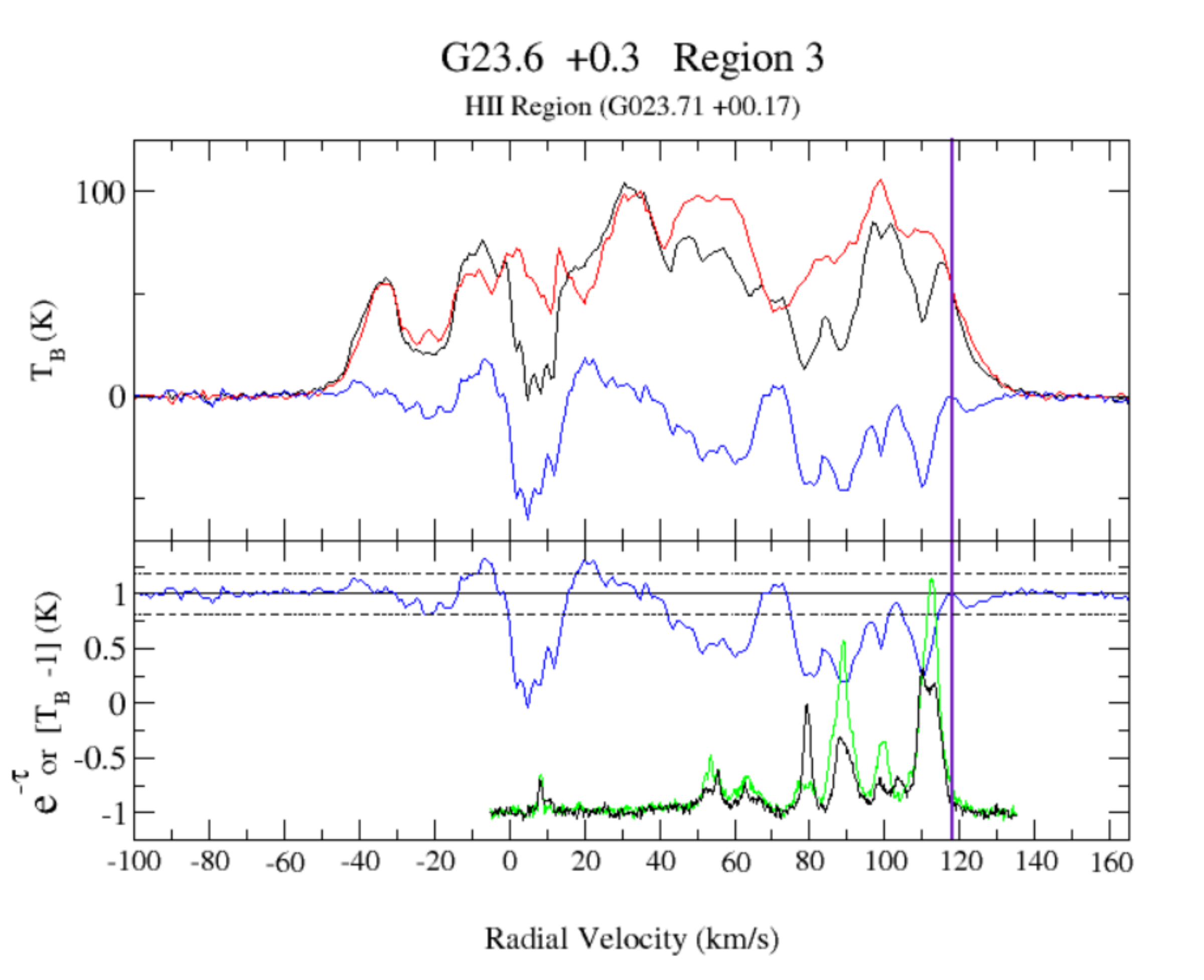}
\caption{G$23.6 +0.3$ spectra (top and middle) and HII region G$023.71+00.17$ (bottom) specrum. Upper half of panel: HI emission spectrum (source: black, background: red and difference: blue). Lower half of panel: HI absorption spectrum (e$^{-\tau}$, blue), $^{13}$CO source (green) \& background (black) spectra (T$_{B}$, offset by subtracting 1 K), $\pm2 \sigma$ noise level of the HI absorption spectrum (dashed line) and tangent point velocity (purple vertical line).}
\label{fig:5}
\end{figure*}

\begin{figure}[ht!]
\centering
\includegraphics[width=\columnwidth]{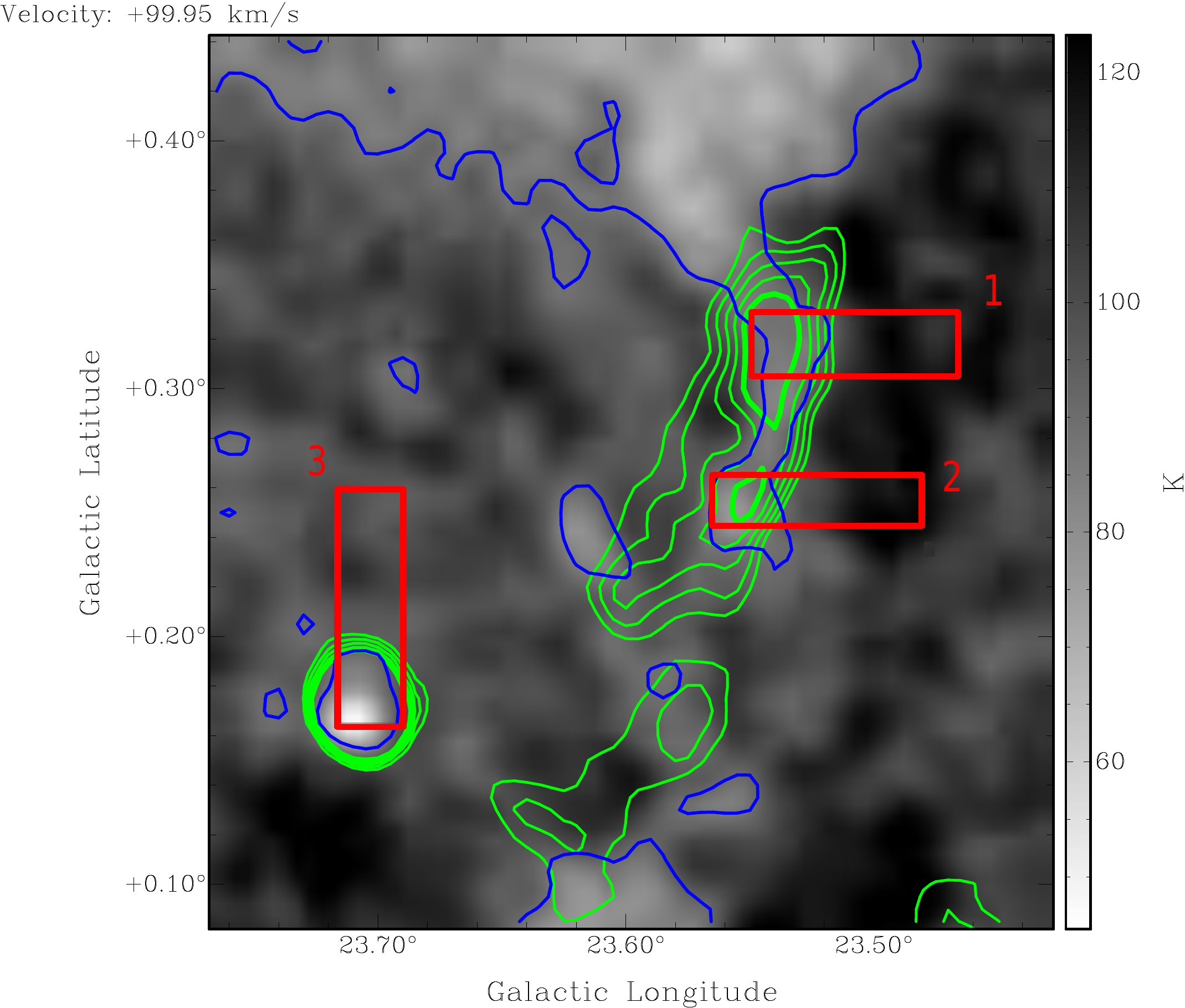}
\caption {G$23.6 +0.3$ HI channel map  +99.95 km s$^{-1} $. HI contour level (Blue): 90 K . Continuum contour levels (green): 32, 34, 36, 38 and 40 K.}
\label{fig:6}
\end{figure}

\indent SNR G$23.6 +0.3$, is a $4^\prime \times 10^\prime$ object located in a complex region, 
in the close proximity to HII regions (e.g. WC89 (G$023.71+00.17$)). 
With a spectral index of 0.34 \citep{Shaver1970}, the oddly elongated shape of the SNR  coincides with 24 $\mu$m emission with no maser association \citep{Goncalves2011}.
  \cite{Shaver1970} quoted a distance of 6.4 kpc to the SNR. 
  The survey \cite{Kilpatrick2016} for broad molecular line regions interacting with molecular clouds places it at a distance of 6.9 kpc.\\
\indent  The regions for  extraction of HI absorption spectra are shown in Figure \ref{fig:4}. 
Regions 1 and 2 were chosen to contain the brightest regions ($\sim30$ K) of the SNR and 
Region 3 was chosen for the HII region G$023.71 +00.17$.  The resulting spectra are shown in Figure \ref{fig:5}.
The spectra of the SNR appear to show absorption up to the tangent point ($\sim118$ km s$^{-1}$). 
However the HI channel maps show that this absorption feature is false: the morphology of the brightest continuum regions do not correlate with the low HI intensity.
Real absorption is seen is at a maximum radial velocity of 99.95 km s$^{-1}$ (Figure \ref{fig:6}), consistent with the spectra and channel maps.  
The $^{13}$CO channel maps show no clear evidence of a molecular cloud interaction.  
Therefore, we place the SNR at the near distance of 5.9 kpc that corresponds to the radial velocity of  99.95 km s$^{-1}$.  \\
\indent The nearby HII region G$23.710 +0.175$ shows absorption up to the tangent point indicating lower limit of distance of 7.6 kpc. 
\cite{Jones2012} presents the radio recombination line velocity of 103.8 km s$^{-1}$ and a distance of $7.78^{+1.03}_{-1.19}$ kpc. 
Using the URC rotation curve we find the distance to the HII region for that velocity to be 9.2 kpc.

\subsection{G$27.4 +0.0$}  \label{G27}

\begin{figure}[ht!]
\centering
\includegraphics[width=\columnwidth]{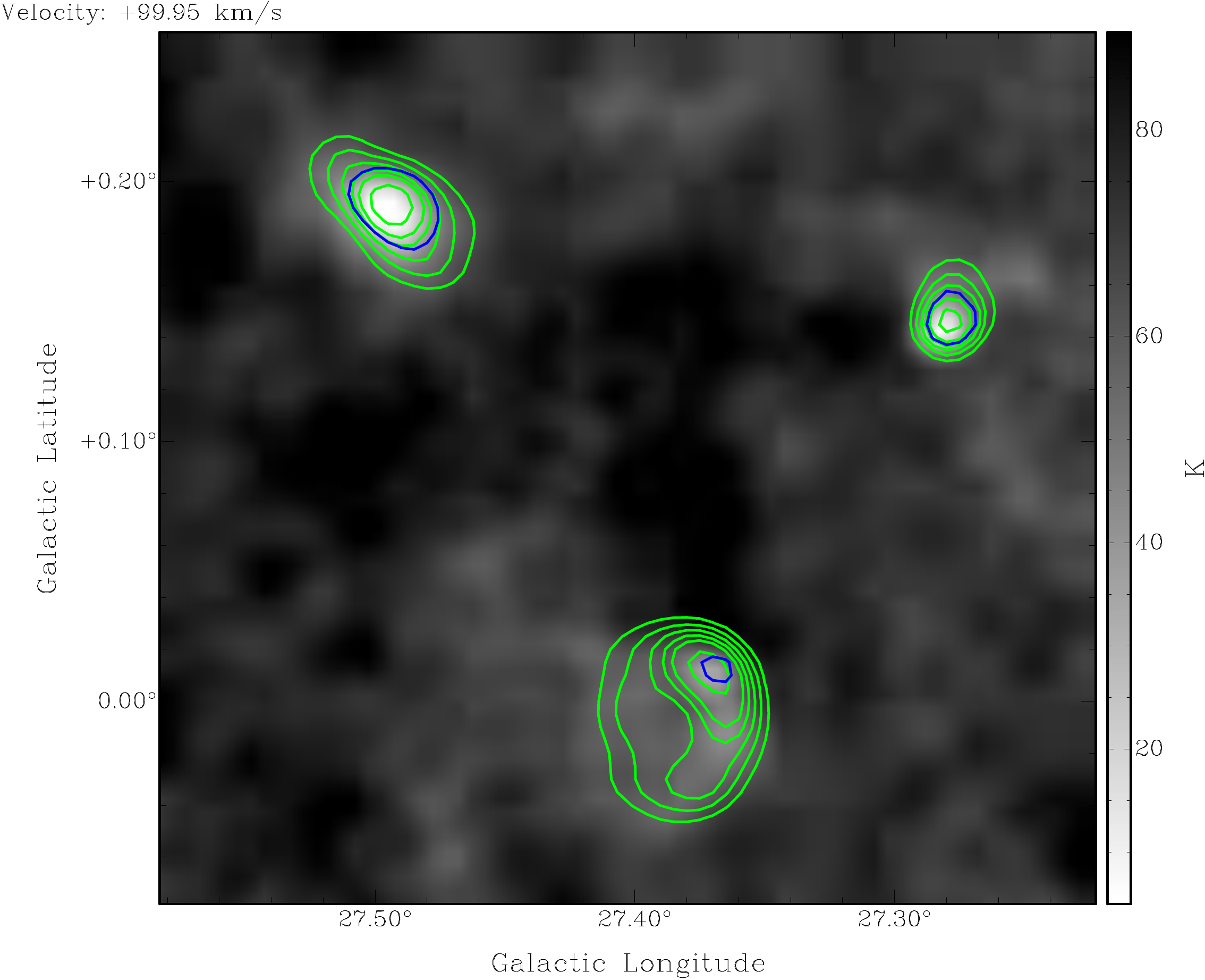}
\includegraphics[width=\columnwidth]{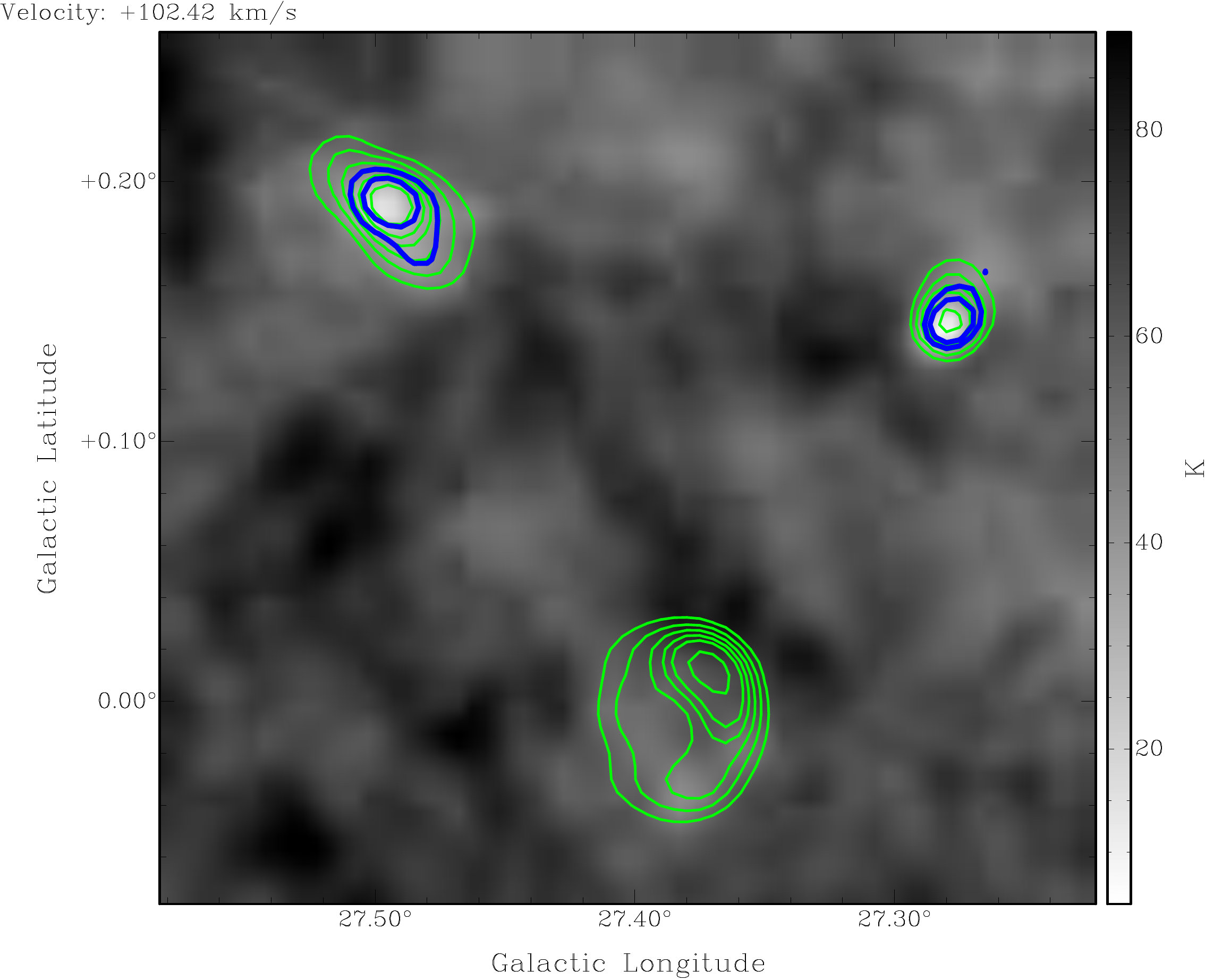}
\caption {G$27.4 +0.0$ HI channel maps  +99.95 and +102.42 km s$^{-1}$. HI contour level (Blue): 40 K for HI channel map +99.95 km s$^{-1}$ and 30 \& 40 K for HI channel map 102.42 km s$^{-1}$. Continuum contour levels (green): 35, 40, 50, 60, 70, 80 and 100 K.}
\label{fig:7}
\end{figure}

\indent Known as 4C-04.71 and Kes 73,  SNR G$27.4 +0.0$ is a shell-type SNR, $ 4.5^\prime \times 5^\prime$ in size. 
It is located near two HII regions, G$27.276 +0.148$ and G$27.491 +0.189$. 
From the X-ray spectrum of the SNR, \cite{Gotthelf1997} give the age to be $\leq 2.2$ kyr. 
The compact central source, 1E 1841-045 associated with G$27.4 +0.0$ was discovered by \cite{Vasisht} as an Anomalous X-ray pulsar (AXP). 
They interpreted its $\sim$11.8 s pulse period and high spin-down rate to obtain a characteristic age of 4.7 kyr. \\
\indent        \cite{Sanbonmatsu1992} presented HI absorption observations for the SNR with estimated distance 6 - 7.5 kpc. 
\cite{TianLeahy2008} revised the distance to be 7.5- 9.8 kpc, leading to an updated age for the SNR of 500-1000 yr and a larger AXP X-ray luminosity.\\
\indent     We constructed HI absorption spectra and verified the absorption features examining the HI channel maps. 
The spectra are essentially the same as given in Figure 2 of \cite{TianLeahy2008}. 
Even though the spectra show absorption features up to the tangent point, the channel maps show that the morphology of HI absorption does not match 
the continuum intensity of the SNR. Thus the absorption near the tangent point is most likely a false feature. 
The maximum radial velocity where  HI absorption for the SNR is seen in the channel maps is  99.95 km s${^{-1}}$ (Figure \ref{fig:7} top panel). 
The conclusion that the SNR is located at the near distance of 5.8 kpc.

We see there is clear absorption for the nearby HII regions up to the tangent point not seen in the SNR (Figure \ref{fig:7} bottom panel).
Thus the SNR and the nearby HII regions are not related.

\subsection{G$33.6 +0.1$}  \label{G33}

\begin{figure}[ht!]
\centering
\includegraphics[width=\columnwidth]{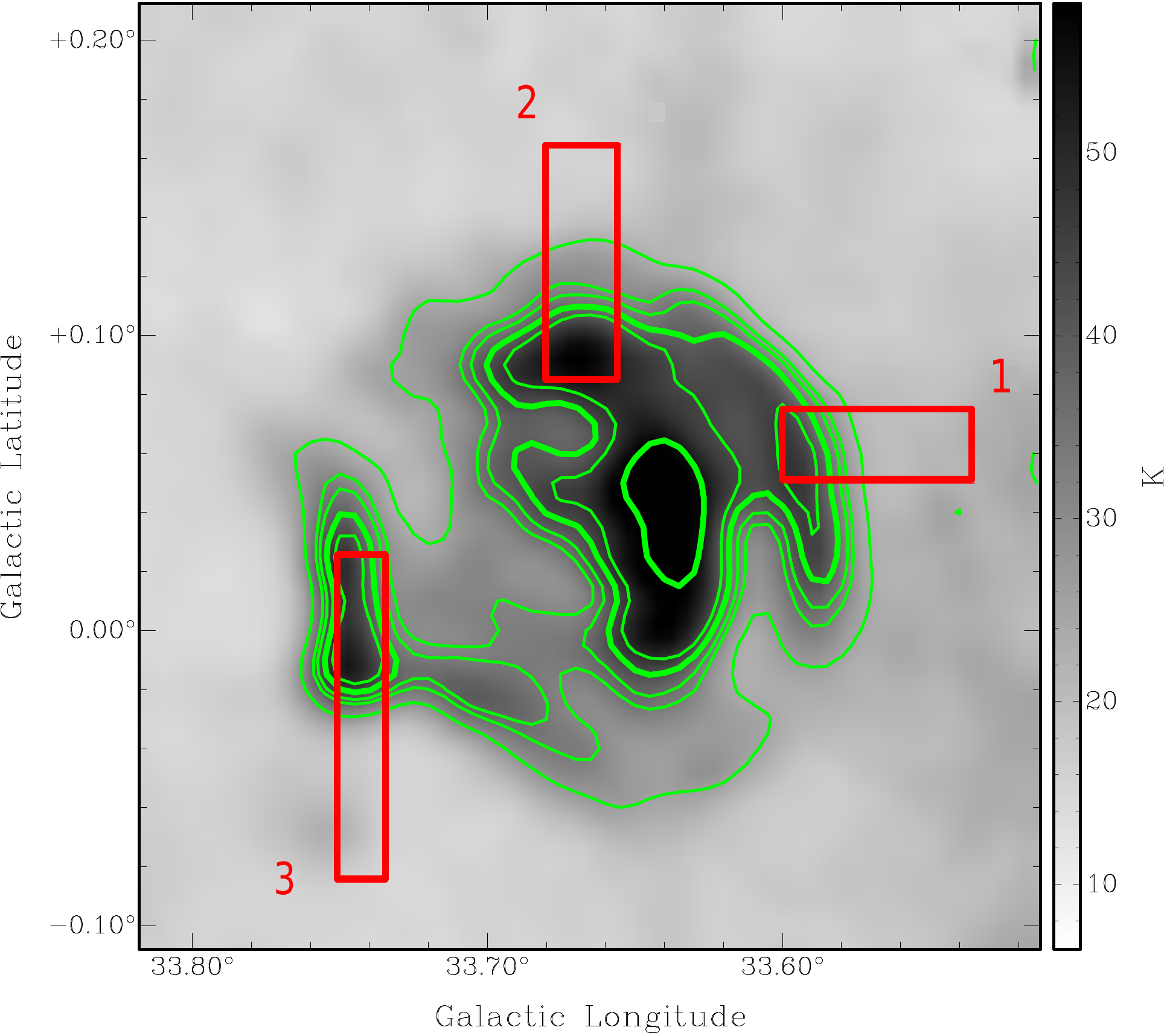}
\caption {SNR G$33.6 +0.1$ 1420 MHz continuum image. Contour levels (green): 25, 32, 35, 40, 45 and 60 K. The red boxes are the regions used to extract HI and $^{13}$CO source and background spectra.}
\label{fig:8}
\end{figure}

\begin{figure*}[ht!]
  \includegraphics[scale=0.35]{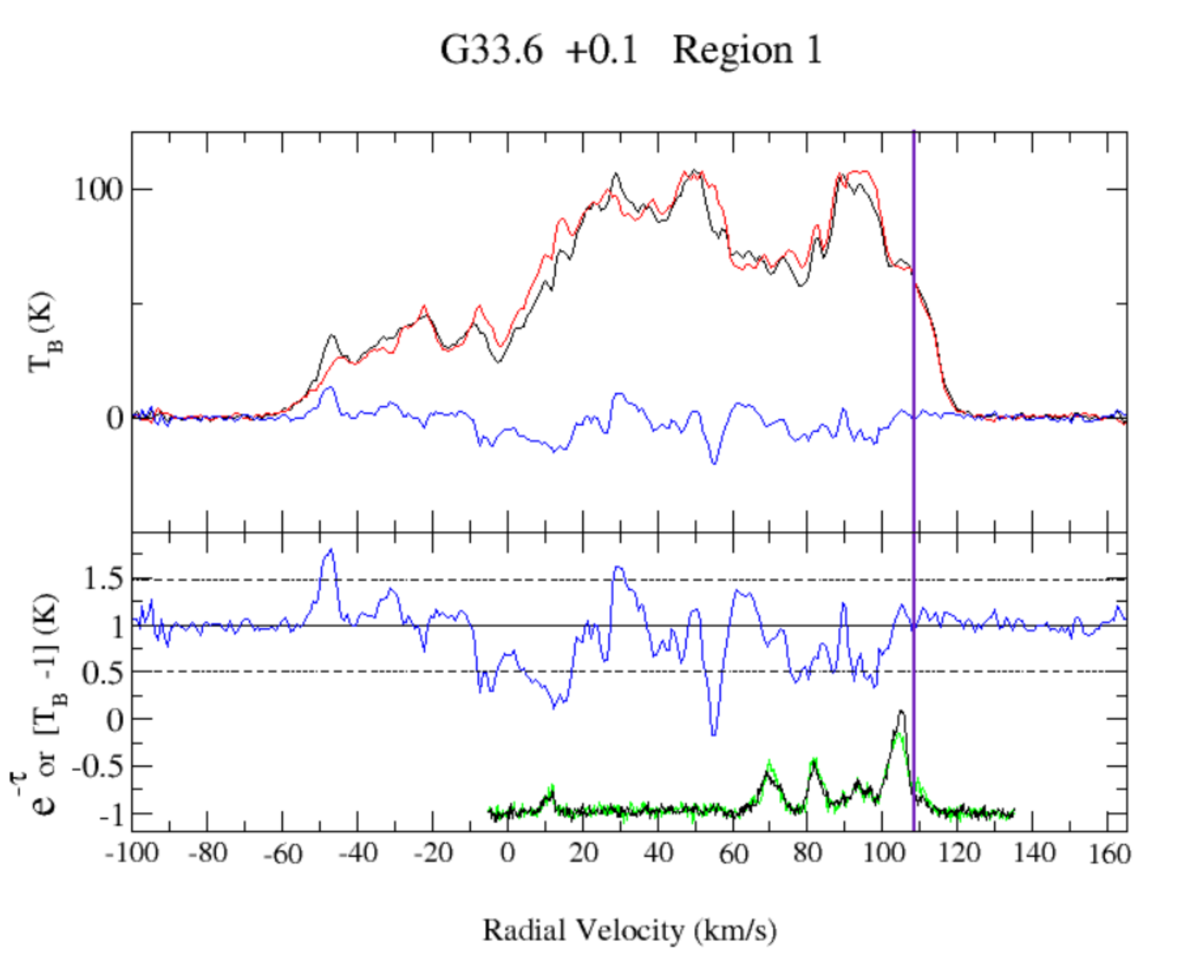}\hfill
  \includegraphics[scale=0.35]{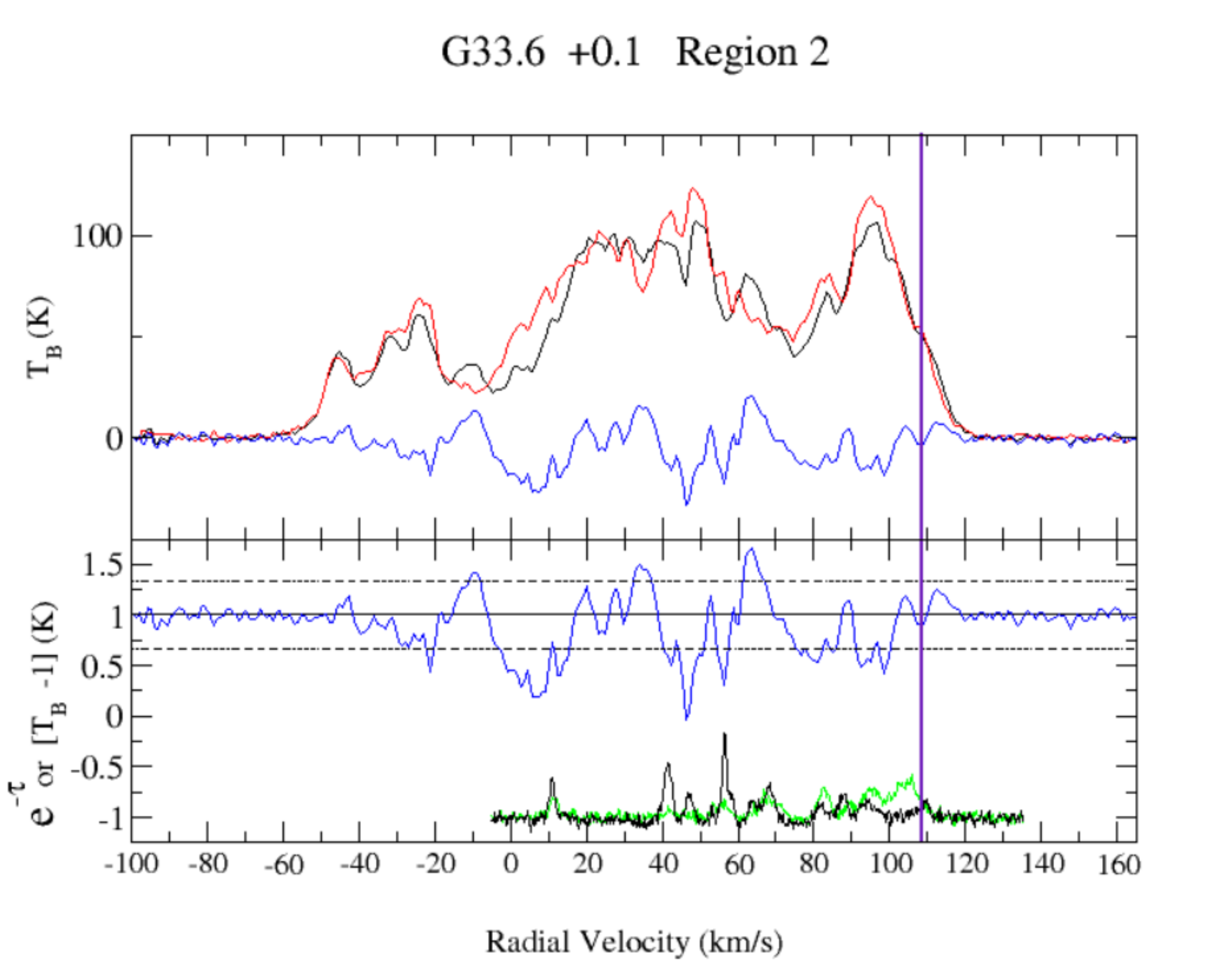}
  \includegraphics[scale=0.35]{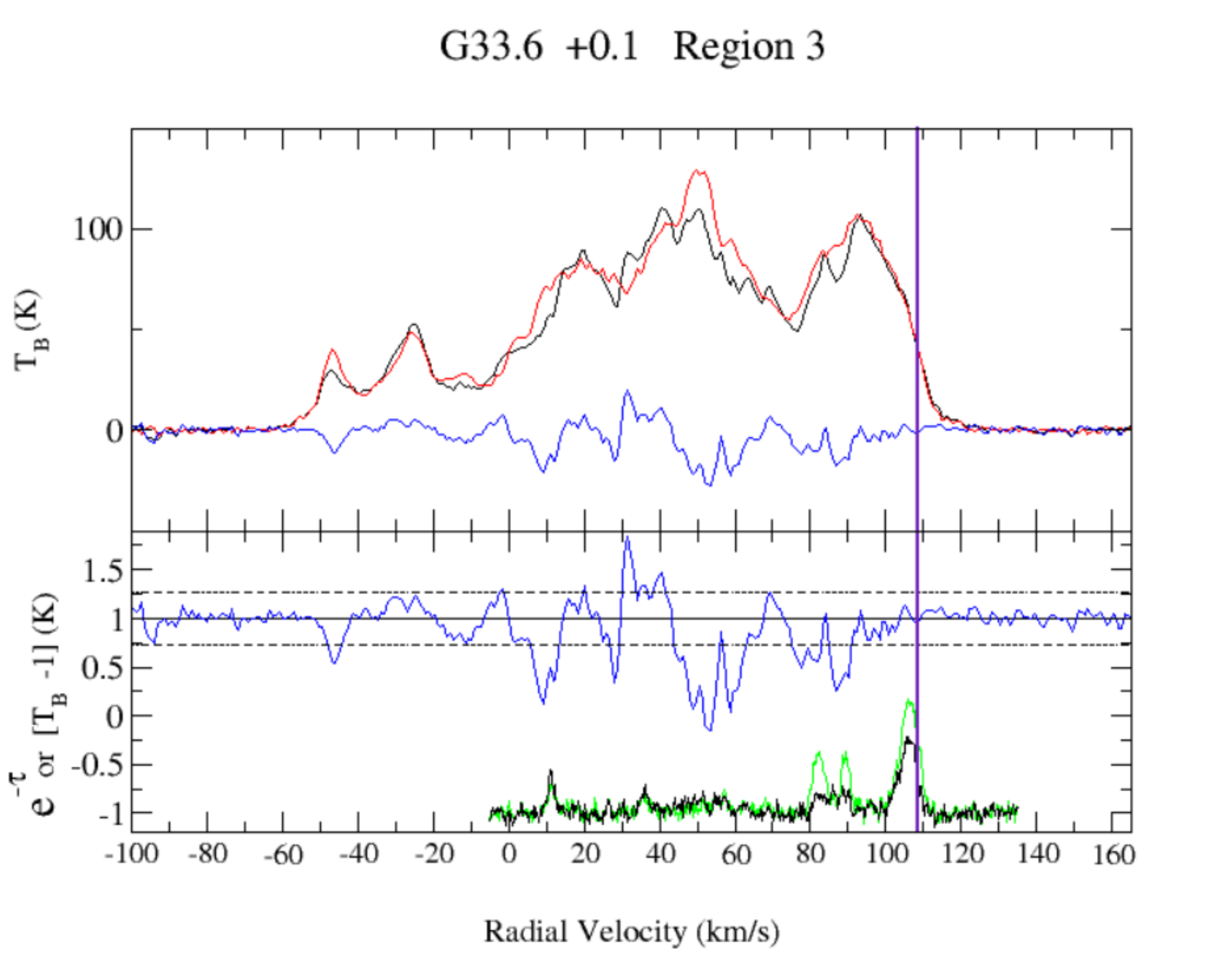}
\caption{G$33.6 +0.1$ spectra. Upper half of panel: HI emission spectrum (source: black, background: red and difference: blue). Lower half of panel: HI absorption spectrum (e$^{-\tau}$, blue), $^{13}$CO source (green) \& background (black) spectra (T$_{B}$, offset by subtracting 1 K), $\pm2 \sigma$ noise level of the HI absorption spectrum (dashed line) and tangent point velocity (purple vertical line).}
\label{fig:9}
\end{figure*}

\begin{figure}[ht!]
\centering
\includegraphics[width=\columnwidth]{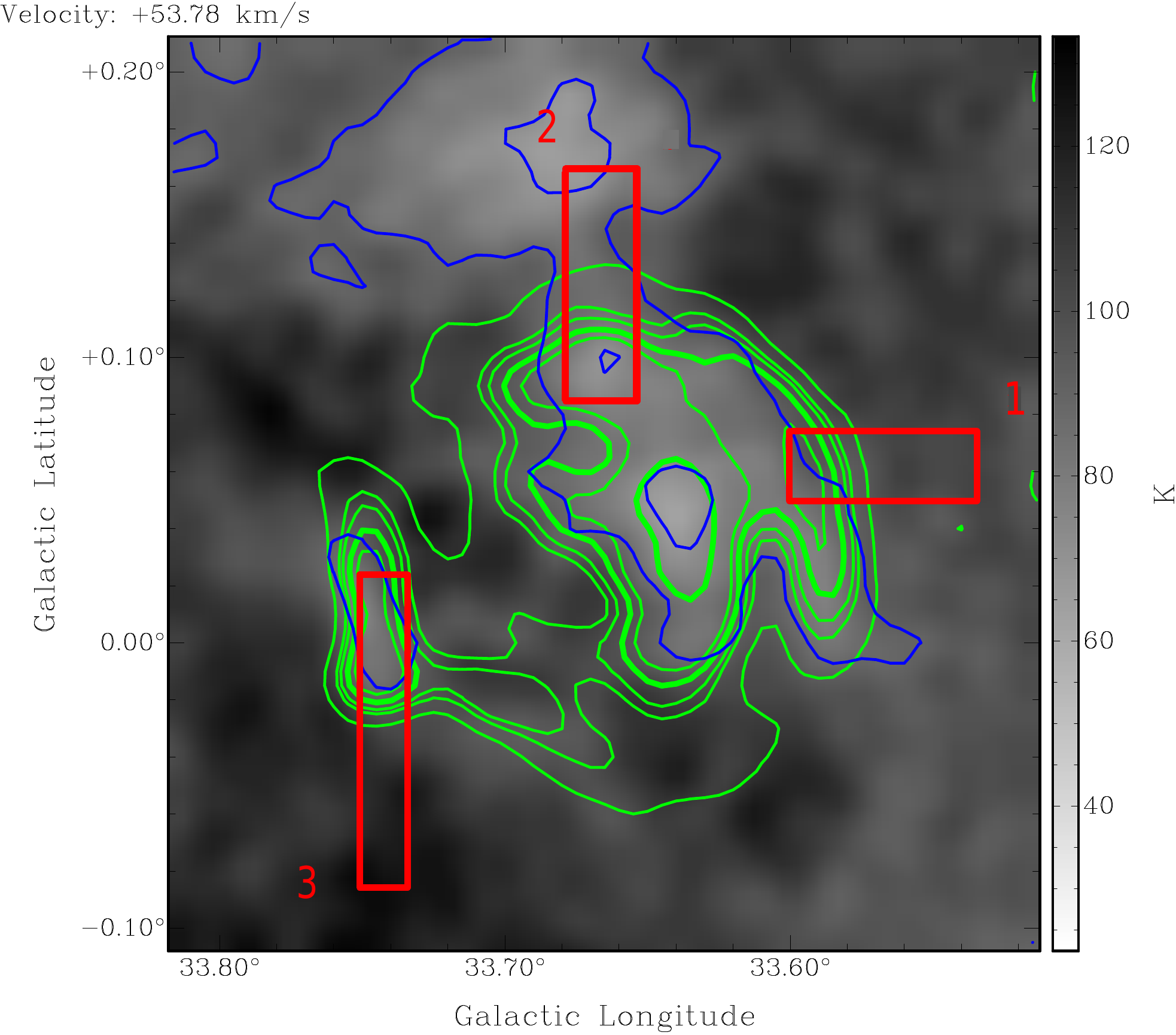}
\includegraphics[width=\columnwidth]{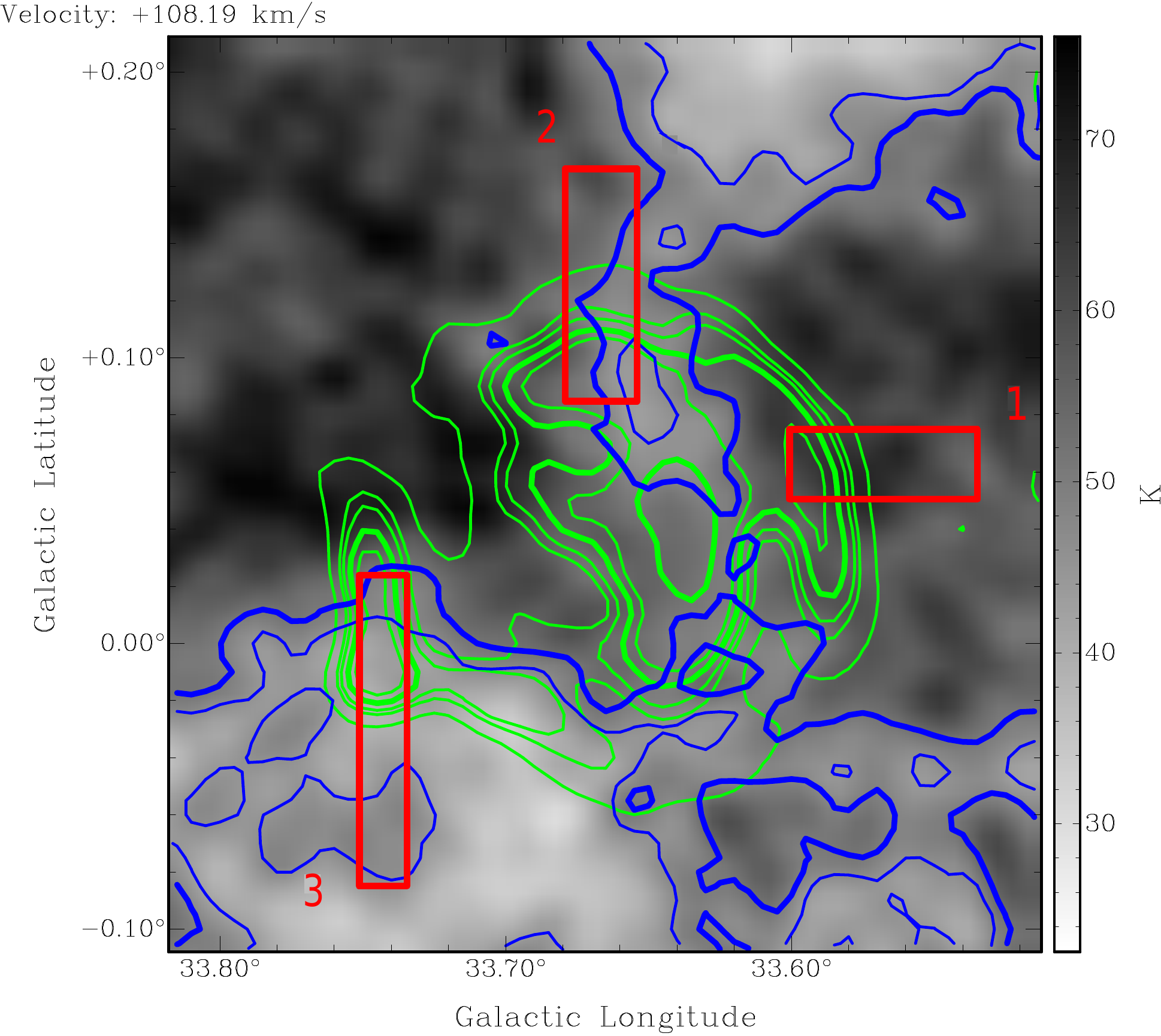}
\caption {G$33.6 +0.1$ HI channel maps  +53.78 and +108.19 km s$^{-1} $. HI contour levels (Blue):  45 and 50 K . Continuum contour levels (green): 25, 32, 35, 40, 45 and 60 K.}
\label{fig:10}
\end{figure}

\indent        G$33.6 +0.1$, is a shell-type SNR located in a complex region. 
Also known as Kes 79, 4C00.70, HC13  and G$33.7 +0.0$, the SNR is $\sim10 ^\prime$ in size and has a bright central region (Figure \ref{fig:8}). 
\cite{Giacani2009} noted the SNR is likely the product of the gravitational collapse of a O9 star evolving near a molecular cloud and within a wind-driven bubble. 
The compact X-ray source CXOU J$185238.6+0.004020$ is located close to the geometric center of the SNR  
and has no association with a radio point source or pulsar wind nebula. 
\cite{Caswell1975} presented an HI absorption spectrum for positive velocities and, because HI  absorption was seen up to the tangent point, suggested a lower limit distance of 7 kpc.
 \cite{Frail1989}, scaling the dispersion measure distance with the ratio of the optical depth integrals, estimated a distance of $10 \pm 2$ kpc for the SNR. 
 \cite{Kilpatrick2016} confirmed a broad line detection and presented a distance of 7.1 kpc consistent with the HI absorption. \\
 \indent  Three spectra extracted from the bright regions (boxes in Figure \ref{fig:8}) are shown in Figure \ref{fig:9}). 
  The left arc and the center, both are bright with a T$_{B} \sim 50 - 70$ K. 
% The brightest region of the SNR is centrally located and excluded in constructing spectra due to the difficulty in defining  reasonable source and background regions. 
The three spectra show several inconsistent features which can be attributed to random clouds that can be seen in the HI channel maps.
The spectra shows no consistent HI absorption at negative velocities. This was verified using the individual HI channel maps. 
 HI absorption is consistently present  in all regions in the spectra and HI channel maps at a velocity of 57.90 km s$^{-1}$ (Figure \ref{fig:10} top panel).
There is no evidence of HI absorption up to the tangent point at $\sim$108 km s$^{-1}$ (Figure \ref{fig:10} bottom panel), even though clear absorption is seen in the nearby objects. %The absorption seen at the higher velocities (near tangent point) has very little correlation between the brightest SNR regions.
From this we derive the distance to the SNR is 3.5 kpc.\\
\indent The examination of $^{13}$CO channel maps show molecular clouds in the vicinity.
 These molecular clouds  do not  morphologically  correlate with G$33.6 +0.1$. The most likely scenario is that these molecular clouds are behind the SNR.

\subsection{G$34.7 -0.4$}

\begin{figure}[ht!]
\centering
\includegraphics[width=\columnwidth]{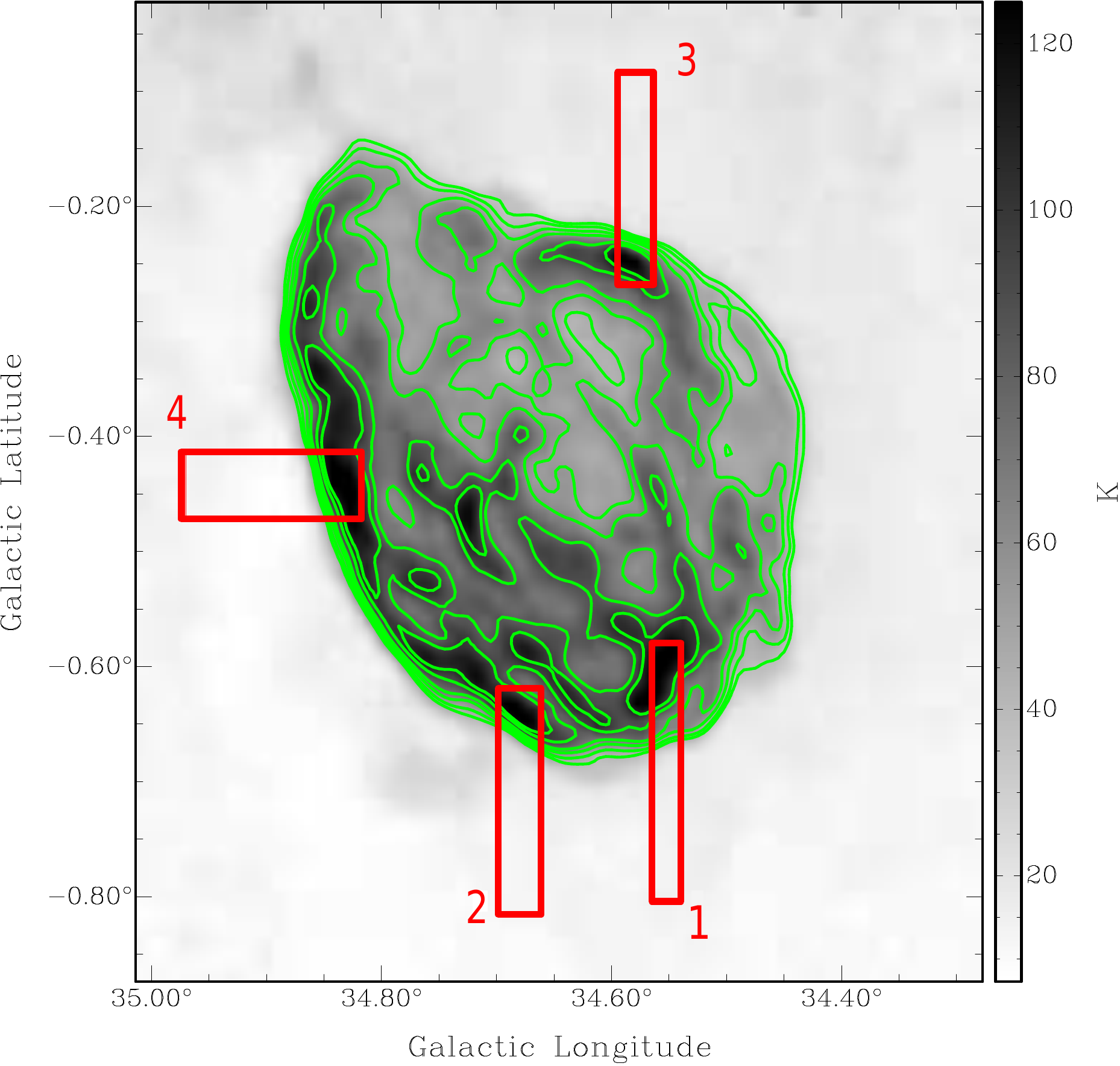}
\caption {SNR G$34.7 -0.4$ 1420 MHz continuum image. Contour levels (green): 38, 45, 55, 65, 85 and 100 K. The red boxes are the regions used to extract HI and $^{13}$CO source and background spectra.}
\label{fig:11}
\end{figure}

\begin{figure}[ht!]
\includegraphics[scale=0.35]{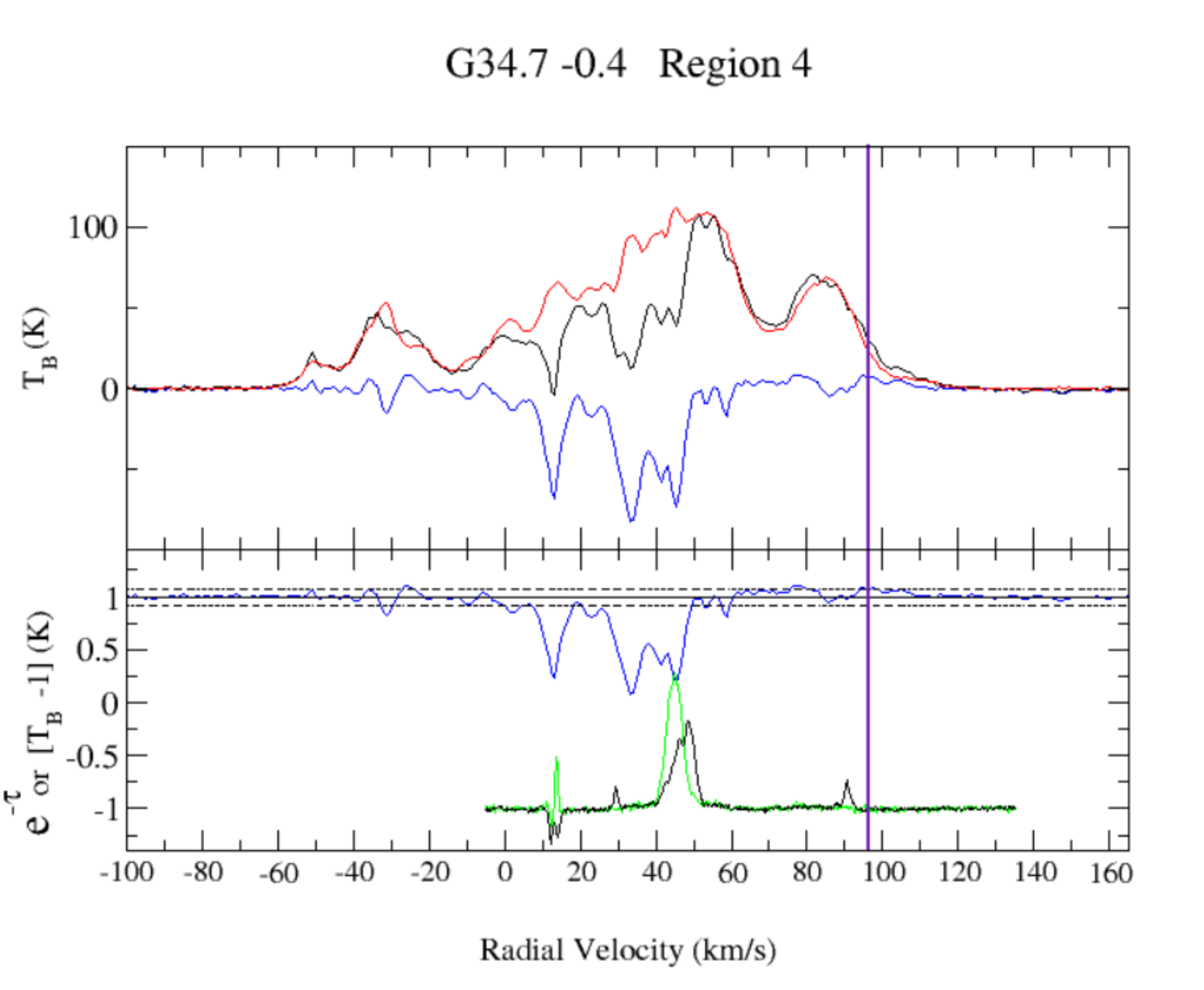}\par 
\caption{G$34.7 -0.4$ spectrum. Upper half of panel: HI emission spectrum (source: black, background: red and difference: blue). Lower half of panel: HI absorption spectrum (e$^{-\tau}$, blue), $^{13}$CO source (green) \& background (black) spectra (T$_{B}$, offset by subtracting 1 K), $\pm2 \sigma$ noise level of the HI absorption spectrum (dashed line) and tangent point velocity (purple vertical line).}
\label{fig:12}
\end{figure}

\begin{figure}[ht!]
\centering
\includegraphics[width=\columnwidth]{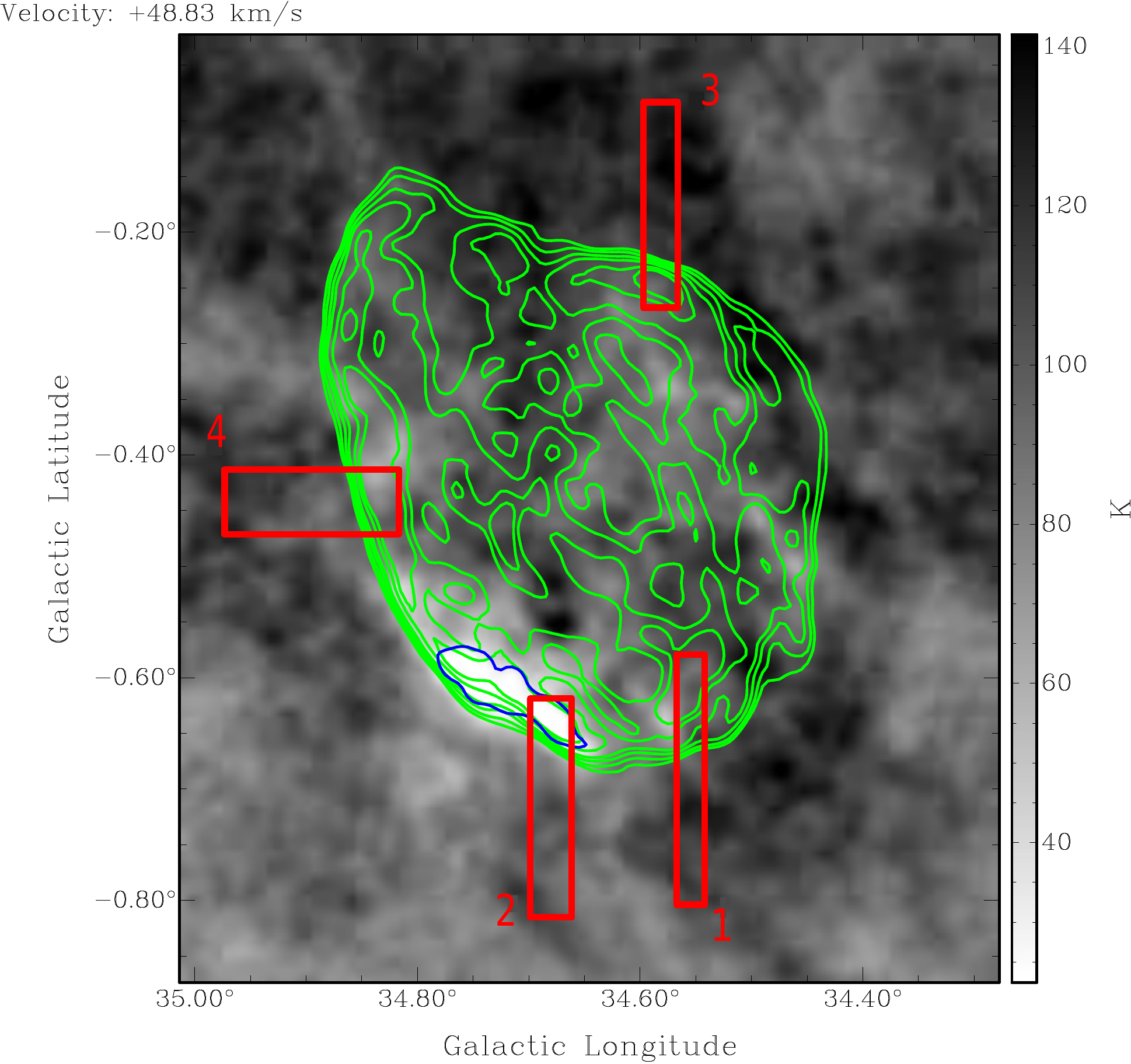}
\caption {G$34.7 -0.4$ HI channel map  +48.83 km s$^{-1} $.  HI contour level (Blue): 30 K. Continuum contour levels (green): 38, 45, 55, 65, 85 and 100 K.}
\label{fig:13}
\end{figure}

 \indent       The SNR G$34.7 -0.4$ is $\sim35^\prime \times 27^\prime $ in size. It is also known as W44, G$34.6 -0.5$  and 3C392. 
 It is a relatively bright remnant with an elongated and distorted shell lying in a complex region, in close proximity to several molecular clouds. 
 \cite{Radha1972} found the distance to be 3 kpc  using HI absorption spectra. This estimate is consistent with OH emission and H${_2}$CO absorption profiles showing 
 strong features at $+45 \pm 5$ km s$^{-1}$ \citep{Ilovaisky1972I}.
 \cite{Caswell1975} found the same distance but with improved HI absorption spectra. 
 \cite{Cox1999} revised the distance estimate to 2.5 - 2.6 kpc using a different Galactic rotation curve and R$_{0}$ = 8.5 kpc. \\
\indent    Figure \ref{fig:11} shows the regions chosen for extraction of HI absorption spectra. 
The western arc of the SNR is the brightest, and all regions were of brightness temperature T$_{B} > 120$ K for the source region. 
All the spectra are consistent with each other, showing no absorption at negative velocities. 
Therefore, we present only the region 4 spectrum  (Figure \ref{fig:12}). 
There is no absorption seen up to the tangent point, indicating the SNR is at the near kinematic distance. 
Continuous strong absorption is seen in the velocity range 0 - $\sim$50 km s$^{-1}$. 
The HI channel map  at V$_{r} = 50.48$ km s$^{-1}$ confirms the absorption along the southern border (Figure \ref{fig:13}), yielding a distance of  3.0 kpc.\\
\indent The $^{13}$CO channel maps  yield no strong evidence of molecular cloud interactions.
There is no morphological evidence that the molecular cloud seen at the velocity range $\sim 42$ to $\sim 50 $ km s$^{-1}$  is interacting with the SNR. 
%The clouds seen at the velocity range $\sim 75 - 80 $ km s$^{-1}$ would be behind G$34.7 -0.4$.\\                

\subsection{G$39.2 -0.3$}  \label{G39}

Also known as 3C396, HC24 and NRAO 593, the SNR G$39.2 -0.3$ is $\sim 8^\prime \times 7^\prime $ in size. 
The radio shell is brighter to the right with a faint tail to the left of the SNR. 
The western shell that is bright in radio and X-ray emission is suggestive of molecular cloud interaction \citep{HewittRho2009}. 
The SNR is located near the HII region NRAO 591. 
\cite{Caswell1975} reported HI absorption present up to the tangent point (7.7 kpc). 
Furthermore, they stated that because the absorption is almost continuous from 60 km s$^{-1}$, the lower limit is 11.3 kpc. 
%They noted a curious feature showing absorption ($ \tau \sim$ 0.4) and enhanced emission at low negative velocities. %Sujiht please check this
\cite{Green1984da} pointed out that the feature at the tangent point of \cite{Caswell1975} may be false and
%Because the absorption is unclear at positive velocities, the minimum  lower limit distance of 11.3 kpc to the SNR may be incorrect. 
 suggested the tangent point distance of 7.7 kpc as the only lower limit distance to the SNR. 
 Later, the revised the lower limit distance to the SNR was 6.6 kpc \cite{Green1989DA}. 
  Based on HI self-absorption, \cite{Su2011} placed the SNR at a distance of 6.2 kpc (tangent point). 

Using the $\Sigma$-D relation, \cite{Patnaik1990} estimated the distance and diameter 7.3 kpc and 16.5 pc respectively for the SNR and 14 kpc for the HII region NRAO 591.
The $\Sigma$-D relation is an observed correlation for SNR with known distances between radio surface brightness $\Sigma$
and physical diameter $D$. Generally, larger SNRs have lower surface brightness. It has quite large scatter \citep{2015Green}, severely limiting its usefulness.

\begin{figure}[ht!]
\centering
\includegraphics[width=\columnwidth]{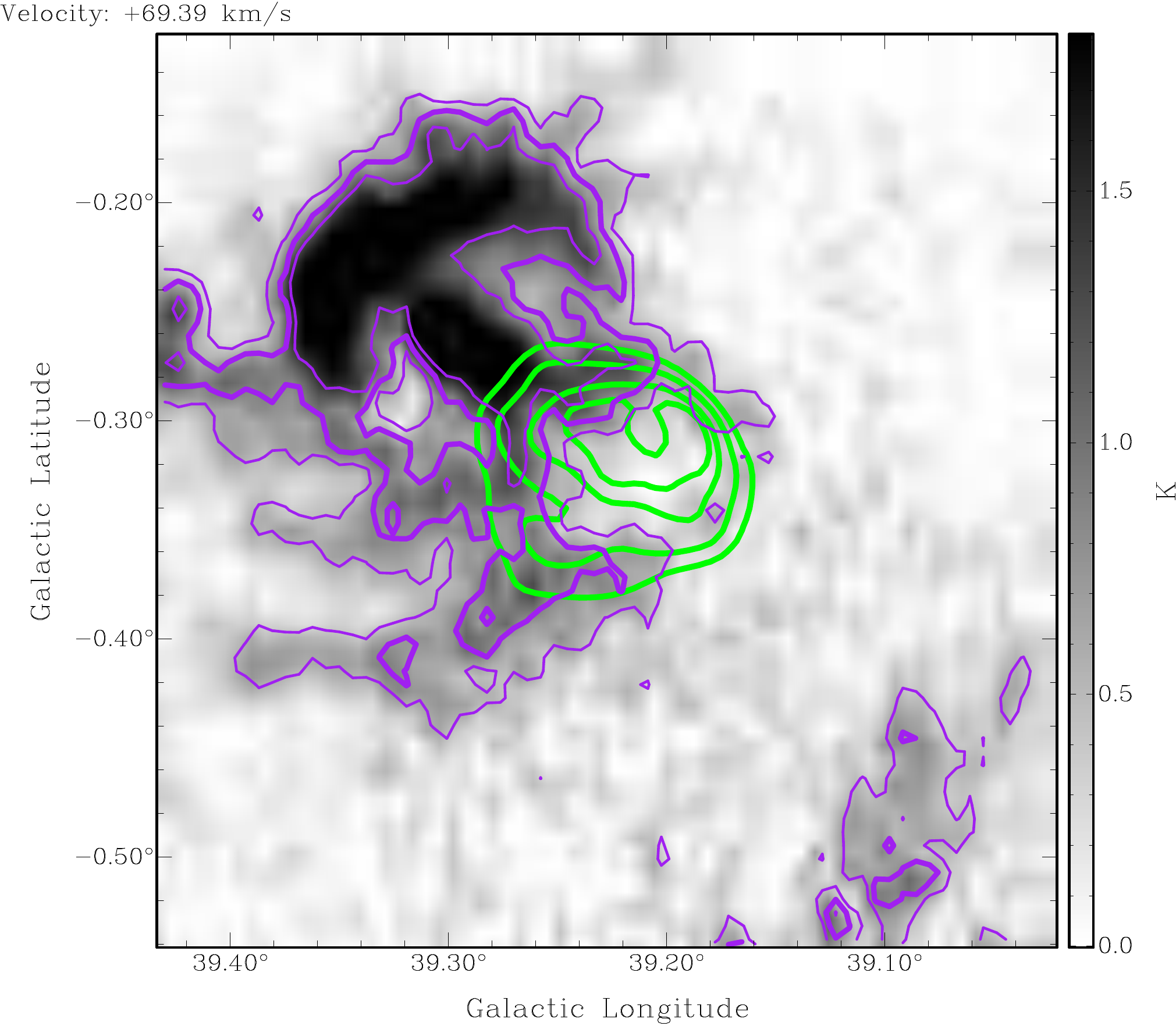}
\includegraphics[width=\columnwidth]{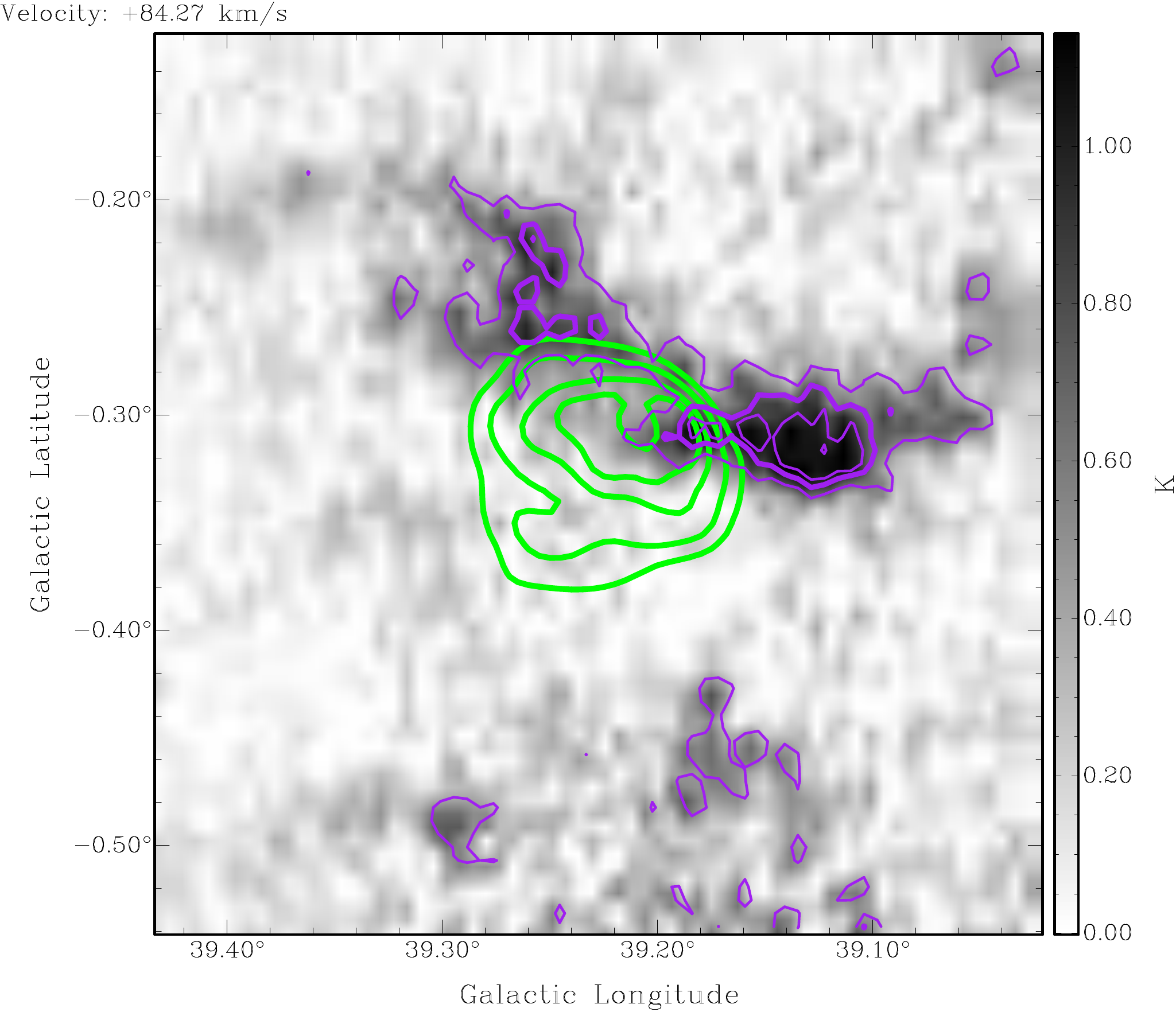}
\caption {G$39.2 -0.3$ $^{13}$CO channel maps  +69.39 and +84.27 km s$^{-1} $. $^{13}$CO contour levels (purple): 0.5, 0.8 and 1.2 K for +69.39 km s$^{-1} $ channel map and 0.5, 0.8 and 1.0 K +84.27 km s$^{-1} $. Continuum contour levels (green) are at 30, 50, 80 and 100 K.}
\label{fig:21}
\end{figure}

\indent We constructed the spectrum for the SNR using the northern bright region (T$_{B}$ = $\sim 120$) and found it to be the same as in \cite{Su2011} (their Figure 7). 
HI absorption is present up to the tangent point yielding a lower limit to the distance of 6.4 kpc. \\
 \indent       The $^{13}$CO channel maps indicate a cavity-like structure between the velocities 67.69 and 70.88 km s$^{-1}$ (Figure \ref{fig:21} top panel). 
 This structure was seen by \cite{Su2011} and is consistent with $^{12}$CO observations. 
 They suggest the cloud seen at $\sim$84 km s$^{-1}$ is more likely to be associated with the SNR. 
 From the $^{13}$CO channel maps, between the velocities $83.63 - 85.12$  km s$^{-1}$  there is a possible molecular cloud interaction  (Figure \ref{fig:21} bottom panel).
However the cavity-like structure at $\sim$69 km s$^{-1}$  spatially correlates better with the SNR. 
Therefore we take the likely distance to G$39.2 -0.3$ as 8.5 kpc  corresponding to the velocity of $69.4$  km s$^{-1}$.

\subsection{G$43.3-0.2$}  \label{G43}

\begin{figure}[ht!]
\centering
\includegraphics[width=\columnwidth]{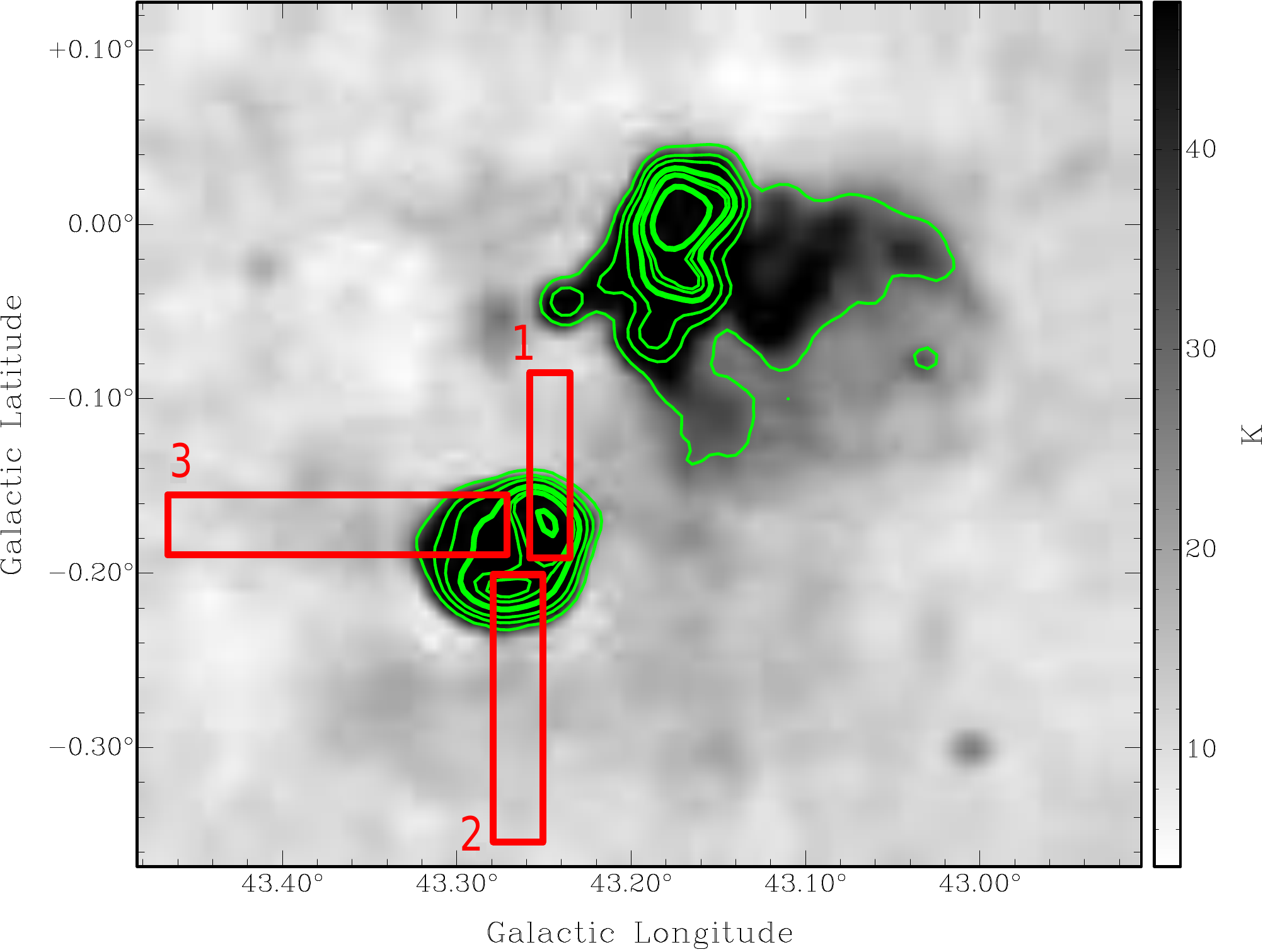}
\caption {SNR G$43.3-0.2$ 1420 MHz continuum image. Contour levels (green): 30, 60, 100, 200, 300, 350, 500 and 600 K. The red boxes are the regions used to extract HI and $^{13}$CO source and background spectra.}
\label{fig:14}
\end{figure}

\begin{figure}[ht!]
\centering
\includegraphics[scale=0.35]{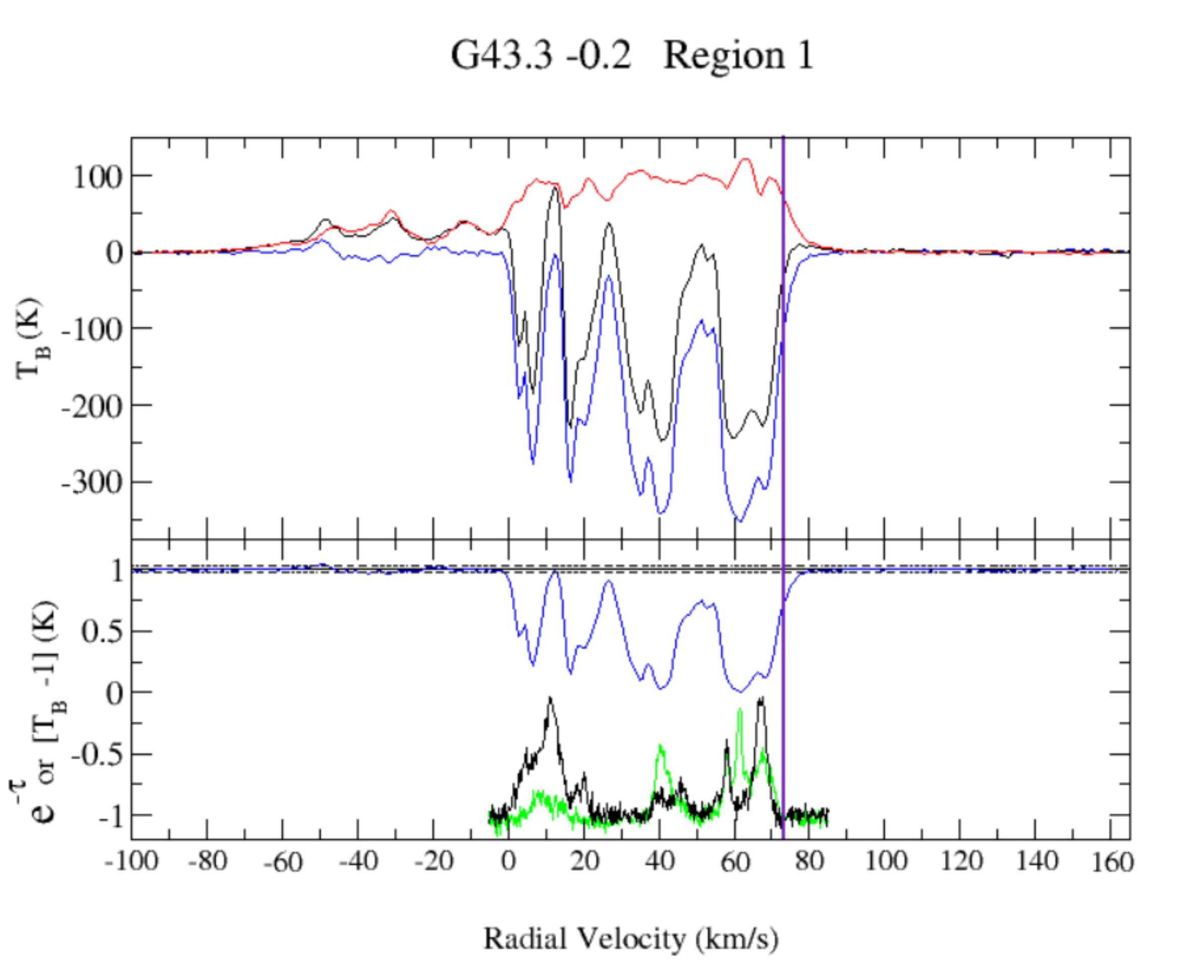}
\caption {G$43.3-0.2$ spectrum. Upper half of panel: HI emission spectrum (source: black, background: red and difference: blue). Lower half of panel: HI absorption spectrum (e$^{-\tau}$, blue), $^{13}$CO source (green) \& background (black) spectra (T$_{B}$, offset by subtracting 1 K), $\pm2 \sigma$ noise level of the HI absorption spectrum (dashed line) and tangent point velocity (purple vertical line).}
\label{fig:15}
\end{figure}

\begin{figure}[ht!]
\centering
\includegraphics[width=\columnwidth]{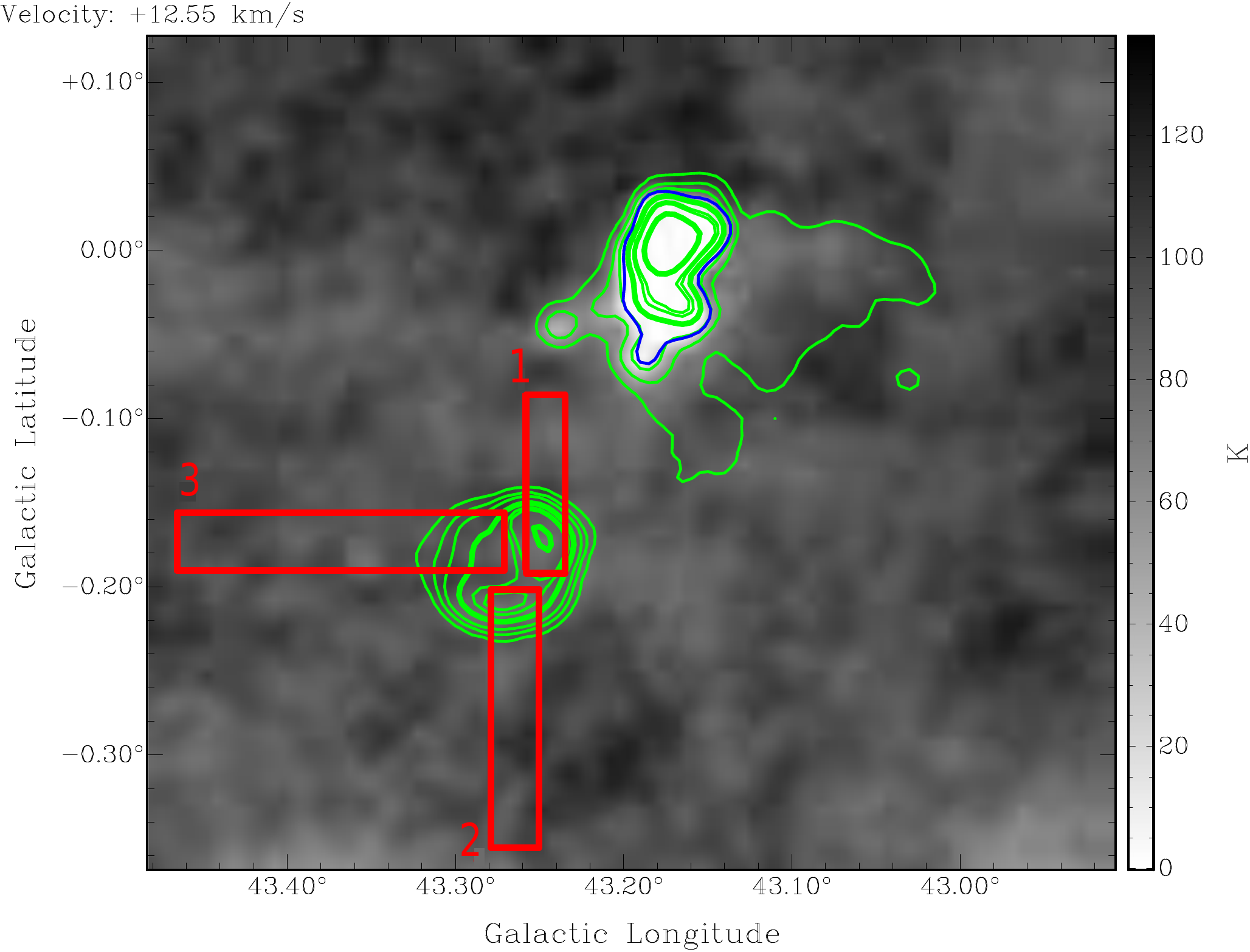}
\includegraphics[width=\columnwidth]{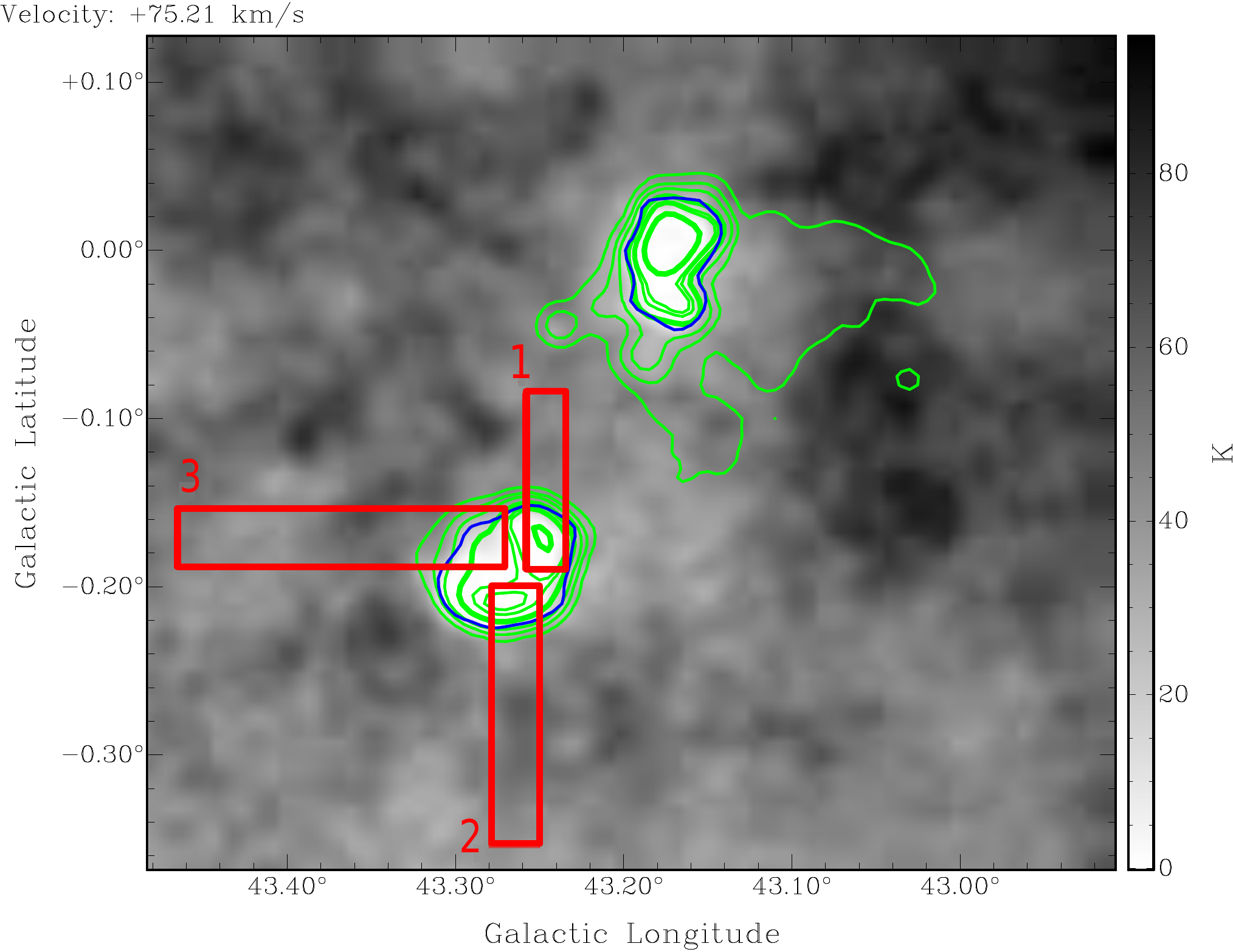}
\caption {G$43.3-0.2$ HI channel maps  at  75.21 and 12.55 and km s$^{-1} $. HI contour levels (Blue):  20 K . Continuum contour levels (green): 30, 60, 100, 200, 300, 350, 500 and 600 K.}
\label{fig:16}
\end{figure}

\indent Known as W49B,  the relatively young SNR G$43.3 -0.2$ is located in the W49 complex which includes one of the most luminous star forming regions in our galaxy. 
The object W49A (G$43.2 -0.0$) near the SNR consists of numerous HII regions \citep{Brogan2001}. 
Early analyses have stated the two objects that are separated by $\sim12^\prime$ are physically associated with each other. 
Analysis of HI  absorption carried out by \cite{LockhartGoss1978} gave the lower and upper limits of distance to G$43.3 -0.2$ as 12.5 kpc  and 14 kpc. 
They noted that the SNR and HII region W49A are possibly associated with each other. % at a distance of 14 kpc. 
\cite{Moffeett1994} revised the distances to 8 and 11 kpc, respectively. 
\cite{Gwinn1992} estimated the distance to W49A to be $11.4 \pm 1.2$ kpc from H$_2$O maser proper motions. \\
\indent Figure \ref{fig:14} shows the continuum image with regions used for spectrum extraction. 
All three spectra are consistent so we present the Region 1 spectrum  in Figure \ref{fig:15}. 
Absorption is seen up to the tangent point. This is verified in the HI channel map (Figure \ref{fig:16} top panel). 
The lower limit of the distance to the SNR is the tangent point distance of 6.0 kpc. 
Examination of the HI channels at negative velocities shows there is no absorption so the SNR is inside the Solar circle.  
The SNR spectrum shows absorption throughout the positive velocities, except with no absorption at velocity $\sim$12 km s$^{-1}$. 
The HI channel map shows absorption in the HII region but not in the SNR  (Figure \ref{fig:16} bottom panel). 
This shows that the SNR is most likely located just this side of V$_r\simeq 12.55$ km s$^{-1}$, corresponding to a distance of 11.3 kpc.  \\
\indent  The HII region W49A shows absorption up to the tangent point. 
However, W49A shows strong absorption at negative velocities, up to V$_r = -22.07$ km s$^{-1}$, beyond the Solar circle at a corresponding distance of 13.7 kpc.
 \cite{Radha1972} places the HII region at $\sim$14 kpc which is consistent with the absorption velocity and distance derived in this study.
  The distance to the HII region W49A, of 13.7 kpc makes it highly unlikely that it's associated with the SNR.                                                                          \\
\indent  Comparison of the $^{13}$CO emission and HI absorption spectra shows that prominent molecular clouds have corresponding HI absorption. 
%This implies that the clouds are in front of the SNR. 
Individual $^{13}$CO channel maps show no evidence of  molecular clouds associated or interacting with  G$43.3 -0.2$.

\subsection{G$46.8-0.3$}  \label{G46}
 
\begin{figure}[ht!]
\centering
\includegraphics[width=\columnwidth]{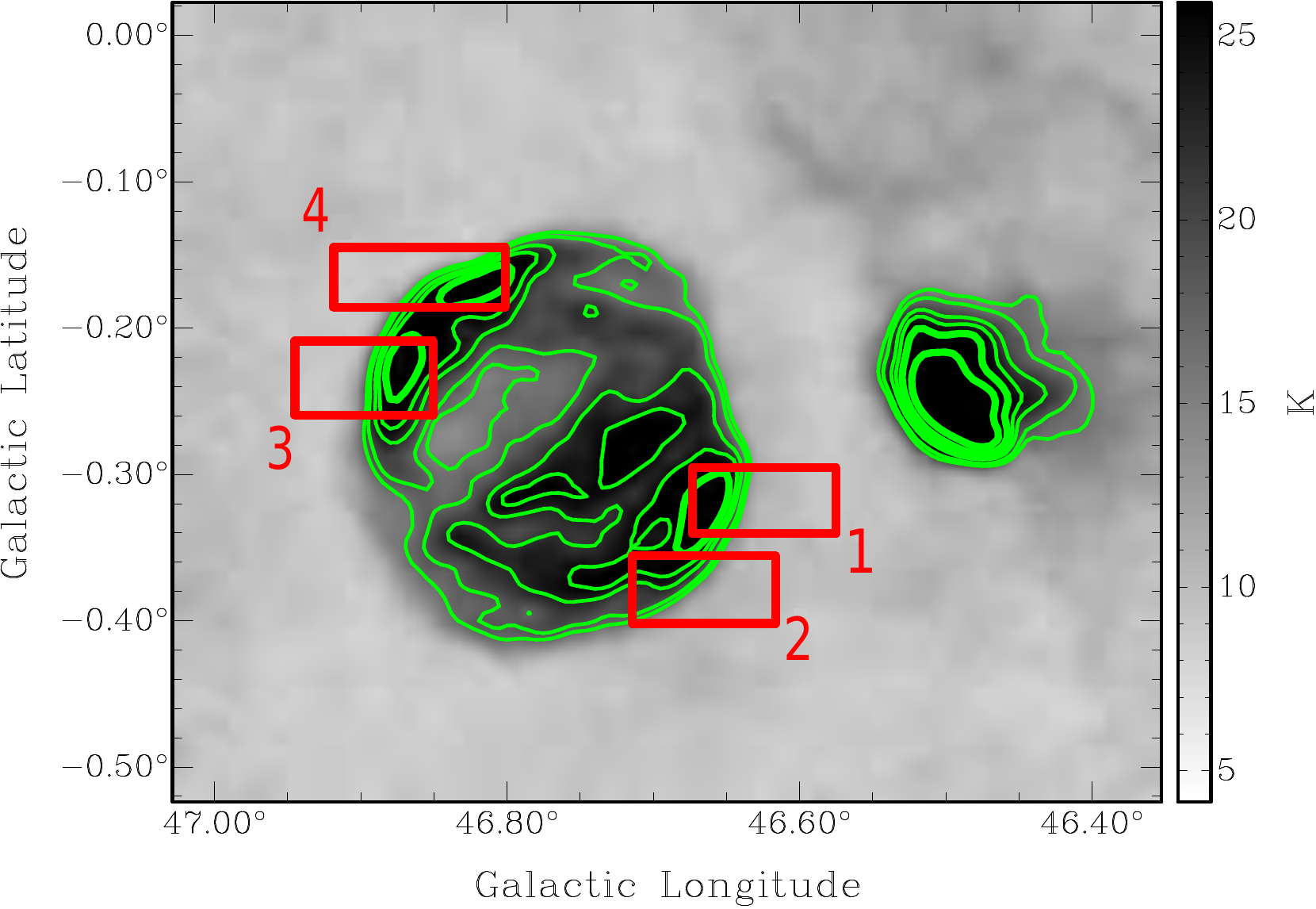}
\caption {SNR G$46.8-0.3$ 1420 MHz continuum image. Contour levels (green) at 16, 18, 22, 25, 30 and 40 K. The red boxes are the regions used to extract HI and $^{13}$CO source and background spectra.}
\label{fig:18}
\end{figure}

\begin{figure}[ht!]
\centering
\includegraphics[scale=0.35]{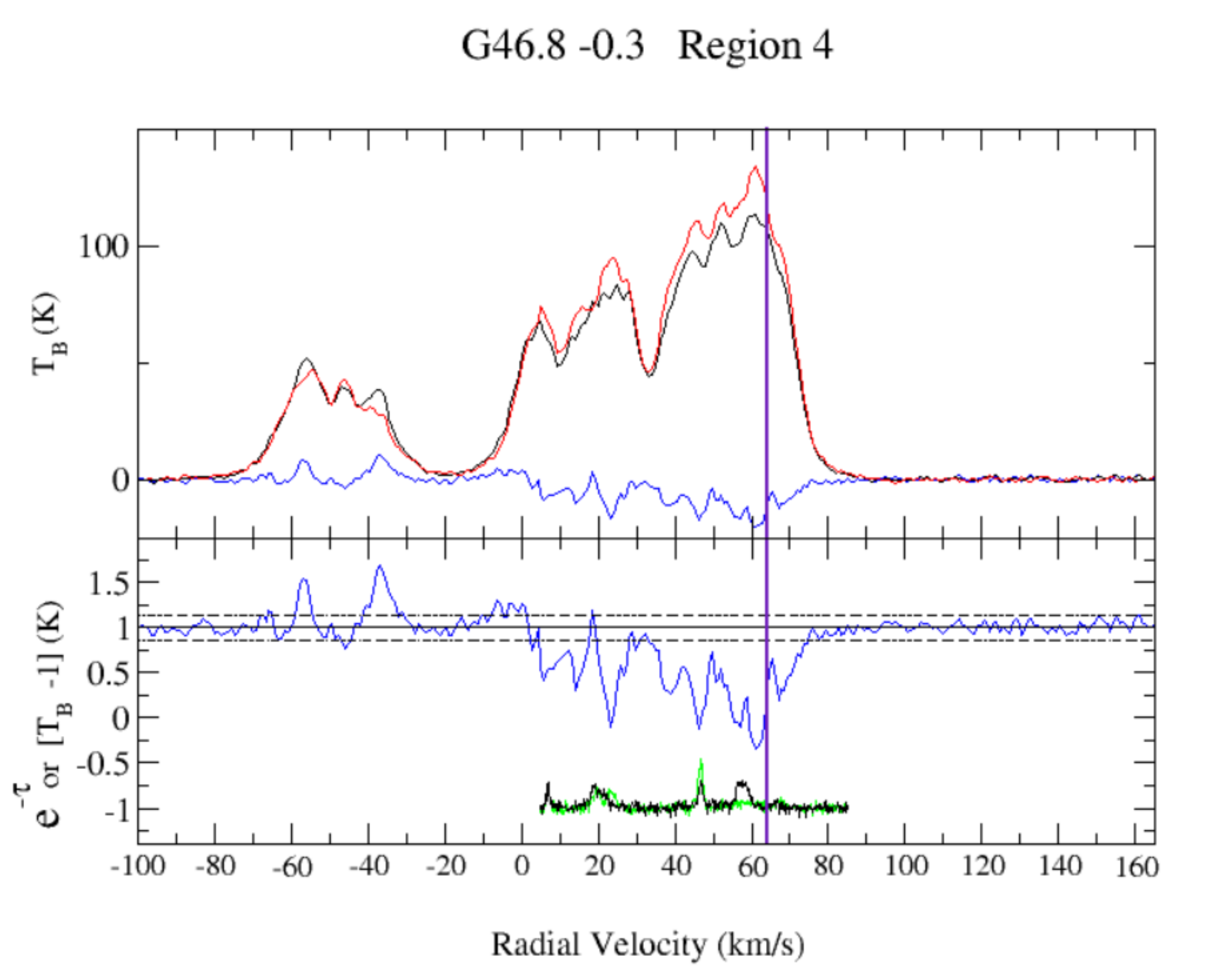}
\caption {G$46.8-0.3$ spectrum. Upper half of panel: HI emission spectrum (source: black, background: red and difference: blue). Lower half of panel: HI absorption spectrum (e$^{-\tau}$, blue), $^{13}$CO source (green) \& background (black) spectra (T$_{B}$, offset by subtracting 1 K), $\pm2 \sigma$ noise level of the HI absorption spectrum (dashed line) and tangent point velocity (purple vertical line).}
\label{fig:20}
\end{figure}

\indent The $16^\prime \times 17^\prime$ SNR G$46.8 -0.3$ (HC30) is located near the HII region G$46.2 -0.2$ (Figure \ref{fig:18}). 
It has an almost complete shell with bright arcs to the north-west and south-east. 
Based on 1.7 and 2.7 GHz observations, \cite{Willis1973} first identified the SNR. 
\cite{Sato1979} suggested a distance between 6.8 and 8.6 kpc from HI absorption and emission features. \\
\indent The region 4 spectrum is shown in Figure \ref{fig:20}. 
HI absorption occurs up to the tangent point velocity at $\sim$65 km s$^{-1}$, thus
the SNR is located  beyond the tangent point distance of 5.7 kpc. 
There are no absorption features at negative velocities giving the upper limit distance at the far side of the solar circle (11.4 kpc).\\
\indent \cite{Quizera2006A} places the HII region at a distance of 7.8 kpc beyond the tangent point. 
The HI channel maps show that the HI absorption for the HII region does not extend to the tangent point ($~64$ km s$^{-1}$) 
and the maximum velocity where the absorption is seen is at $\sim57$ km s$^{-1}$. 
This velocity is consistent with the radio recombination line velocity (RRL) of 57.09 km s$^{-1}$ \citep{Quizera2006A}. We place the HII region at the near-distance of  4.1 kpc.      

\subsection{G$49.2-0.7$}  \label{G49}

Known as W51C, G$49.2 -0.7$ is located in the W51 complex, a star forming complex that consist of HII regions W51A, W51B and the SNR W51C (\cite{TianLeahy2013} Figure 1). 
%The extent and structure of the SNR has been difficult to determine. 
The SNR is a partial shell that extends on to the complex W51B.
Magnetohydrodynamic simulations were carried out to disentangle the SNR from its surrounding emission \citep{2017Zhang}.
 \cite{Sato1973} used HI absorption to estimate the distance of W51C as 4.1 kpc. 
%However, the HI survey that was used couldn't resolve the SNR with the thermal components.  
\cite{Koo1995} associated G$49.2 -0.7$ with a high-velocity molecular cloud that placed the SNR at $\sim$6 kpc, near the the tangent point distance, 
but points out the cloud may extend over 1.5 kpc along the line of sight making it unclear where the SNR is located. 
\cite{TianLeahy2013} reported a distance of 4.3 kpc to the SNR  using HI absorption spectra. 

Our HI absorption spectra are essentially the same as given in  Figure 2 of \cite{TianLeahy2013}. 
The rotation curve (URC with \cite{Reid2014} parameters) yields a tangent point velocity of 57.8 km s$^{-1}$, inconsistent with the emission spectrum towards the SNR. 
Our fit of the observed HI profile including a Gaussian velocity dispersion yields a tangent point velocity of 65.6 km s$^{-1}$. 
Absorption is present up to the tangent point, giving a lower limit distance to the SNR of the tangent point distance 5.4 kpc. 
\cite{Koo1997} noted the molecular cloud associated with the SNR is at $62 \pm 2$ km s$^{-1}$. 
The nearby HII regions \citep{Quizera2006A} G$48.930-0.28$, G$48.997-0.29$, G$49.204-0.34$ and G$49.384-0.30$ have RRL velocities of 65.86, 66.86, 67.79 and 67.38 km s$^{-1}$ respectively. 
\cite{Quizera2006B} places the HII regions at $\sim5.6$ kpc (tangent point with R$_{0} = 8.5 kpc$). 
We conclude that the SNR and the HII regions are likely associated with each other and place the SNR and the HII regions at a distance of 5.4 kpc.  

\subsection{G$54.1+0.3$}  \label{G54}

\indent The $\sim 12^\prime$ SNR is a possible composite type remnant around the Crab-like pulsar wind nebula (PWN) G$54.1 +0.3$ and located near the HII region G$54.09 -0.06$, and compact sources G$53.83 -0.06$ and G$54.1 +0.1$ \citep{LeahyTianWang2008}. 
The relatively bright pulsar wind nebula core is  $\sim 3^\prime$ across. 
\cite{LU2002} estimated the distance to the SNR as $\sim$5 kpc, based on X-ray observations, with absorption column density about half the Galactic absorption. \cite{LeahyTianWang2008} gave a distance of $4.5 - 9 $ kpc to the SNR, using HI absorption spectra. 
They suggested the morphological association of the PWN with a molecular cloud at $53$ km s$^{-1}$ yields the likely distance to the SNR of $6.2^{+1.0}_{-0.6}$ kpc. \\     
\indent Our HI spectra for the PWN are the same as  Figure 2  in \cite{LeahyTianWang2008}. 
The URC rotation curve gives a tangent point velocity of 45.5 km s$^{-1}$,  inconsistent with the emission spectrum towards the SNR. 
Our fit of the observed HI profile including a Gaussian velocity dispersion yields a tangent point velocity of 52.8 km s$^{-1}$, very different than that
assumed by \cite{LeahyTianWang2008} . 
The associated $^{13}$CO molecular cloud at 53.66 km s$^{-1}$ (\cite{LeahyTianWang2008} Figure 3) places the SNR at the tangent point. 
Therefore, the distance to the SNR is 4.9 kpc.

\section{Discussion and Summary} \label{sec:dissummary}

We have analyzed HI 21 cm line observations and $^{13}$CO  line observations of 21 supernova remnants (SNRs)
which are located in the sky area of the VGPS survey. 
The Galactic rotation curve we use for that area of the Galaxy is the URC rotation curve with \cite{Reid2014} parameters.
We examined 1420MHz continuum emission, HI emission and absorption spectra,  $^{13}$CO emission spectra and the HI line $^{13}$CO  line channel maps.
We obtain new observational evidence for 10 SNRs and thus revise their distances to be different than previously published values.
For the other 11 SNRs, we confirm the kinematic velocities but use an updated error analysis and the updated rotation curve to revise distances. 
 The uncertainties in  $V_r$ and the resulting distance are discussed in \cite{2017Ranasinghe}.

Table \ref{tab:table1} presents our results for the 21 SNRs. 
The literature distances are presented with a note on the method, whether by HI absorption, association with molecular cloud or by the $\Sigma-D$ relation. 
 The various assumed values of radial velocity for the SNR and parameters of the rotation curve ($R_0$ and $V_0$) used in previous distances are listed in columns 4 to 6. 
 Our current most likely radial velocity, $V_r$ is listed in column 8.
 Column 9 notes whether the SNR is at the near side or far side of the tangent point, or at the tangent point. 
 The new distance is listed in column 10.
Column 11 notes whether there is an association of the SNR with a molecular cloud.

We note that previous work has used a wide range of Galactic rotation curves, usually assuming $V(R)$ is constant. In some cases the assumed rotation
curve is not quoted. The mean and standard deviation for the quoted $R_{0}$ values
is 8.25 kpc and 0.56 kpc, and for the $V-{0}$ values is 223 km s$ ^{-1}$ and 12 km s$ ^{-1}$. This range in $R_{0}$ and $V_{0}$ can by itself can lead to a range of distances.
For many cases our radial velocities are similar to those in the literature.  They agree within 5 km s$ ^{-1}$ for 9 cases, are very different for 3 cases, and for the remaining cases there
is no well defined value in the literature.

Next we discuss the improvement to the distances to the SNRs, in addition to the fact that we use a consistent rotation curve here, rather than the different rotation curves
used for each SNR in the literature.

The distance has significantly changed (more than 2$\sigma$), based on our consistently derived errors, for 9
SNRs G$18.1-0.1$, G$18.8 +0.3$, G$20.0 -0.2$,  G$23.3 -0.3$, G$23.6 +0.3$, G$27.4 +0.0$, G$33.6 +0.1$, G$39.2 -0.3$ and G$54.1  +0.3$.
For 8 SNRs the old and new distances agree within 2$\sigma$: SNRs
G$18.6 -0.2$, G$21.5 -0.9$, G$22.7 -0.2$, G$32.8 -0.1$, G$34.7 -0.4$, G$35.6 -0.4$, G$41.1 -0.3$  and G$46.8 -0.3$. 
For 4 SNRs the previous distance was very poorly known G$21.8 -0.6$, G$29.7 -0.3$, G$43.3 -0.2$  and G$49.2 -0.7$. 
Including the first two groups of 21 SNRs, the mean change in distance is 1.5 kpc. This is quite significant,
considering the distance range is 3 to 13.8 kpc with mean 6.4 kpc. I.e. on average the distance improvement is $23.4\%$.
The largest change in distance is for G$20.2-0.2$ which went from an old estimate of 4.5 kpc to a new value of 11.2 kpc.
The improved distances have a significant effect on the physical interpretation of supernova remnants. 
For example, we use the Sedov model equations as given in \citet{1972Cox}. 
The inferred shock radius $R_s$ of an SNR depends linearly on distance, $d$.
As described in \citet{2017LeahyWilliams}, the ISM density $n_{0}$ depends on the
emission measure, $EM$, as $\sqrt{EM}$, and $EM$ depends on distance as $d^2$. Explosion energy $E_{0}$
depends on $EM$, thus on $d^2$. 
For the Sedov model, the important quantity is $E_{0}/n_{0}$, which depends linearly on distance. 
SNR age in the Sedov model depends on $R_s^{5/2}  (n_{0}/E_{0})^{1/2}$, thus depends on $d^2$.
For more sophisticated models, the scaling with distances are not as simple as this, but the change in results is similar.

%5 cases have the old values the same as ours within our uncertainties. Another 3 cases have distances the same within 2 $\sigma$ errors. 
%For 7 cases we have distances more than 3 $\sigma$ different than old values. For the remaining cases 
 
In summary, we have obtained distances to a significant number of SNRs using a consistent method and rotation curve. Follow-up work will investigate the implications for the ages and evolutionary states of these SNRs by incorporating other data.
For example, with X-ray spectra the shock temperature and the emission measure can be determined.
These can be used to estimate physical properties of SNRs using explosion models. % in similar manner to the study by  \citep{2017Leahy}.
Easy-to-use SNR explosion models can be based on analytical fits to numerical explosion calculations. 
Such models have been presented by \cite{2017LeahyWilliams}, based on the \cite{1999TM} models.
These models have been applied to the set of LMC SNRS by \cite{2017Leahy} to obtain important properties of the
SNR population, including finding a log-normal distribution of explosion energies, the LMC SNR birthrate, and the distribution of
ISM densities around SNRs in the LMC.
Similar methods will be applied to learn about the properties of Galactic SNRs.

\begin{deluxetable*}{cccccccccccc} \label{tab:table1}
\tabletypesize{\scriptsize }
\tablecaption{Distances to supernova remnants}

\tablenum{1}

\tablehead{ \colhead{\#} & \colhead{Source} & \colhead{} & \colhead{} & \colhead{Literature} & \colhead{} & \colhead{}  & { } & \colhead{} & \colhead{New results} & \colhead{} & \colhead{} \\  \cline{3-7}
\cline{9-11}
\colhead{} & \colhead{} & \colhead{Dist\tablenotemark{a}} & \colhead{V$_{r}$} & \colhead{R$_{0}$} & \colhead{V$_{0}$}& \colhead{Refs} & &\colhead{V$_{r} $} & \colhead{KDAR\tablenotemark{b}} & \colhead{Dist} & \colhead{$^{13}$CO\tablenotemark{c}} \\
\colhead{} & \colhead{} & \colhead{(kpc)} & \colhead{(km s$^{-1}$)} & \colhead{(kpc)} & \colhead{(km s$^{-1}$)}& \colhead{ } & & \colhead{(km s$^{-1}$)} & \colhead{} & \colhead{(kpc)} & \colhead{} }

%% All data must appear between the \startdata and \enddata commands
\startdata
01 & G18.1 -0.1  & $5.6$\tablenotemark{H} & 100 & 8.5 & 210 & 1 & & 103.74  & N   & $6.4 \pm 0.2$ & Possible\\ 
02 & G18.6 -0.2  & $4.6 \pm 0.6$\tablenotemark{H} &  62 & 8.5 & 220 &  2 &  &  62.84  & N   & $4.4 \pm 0.2$& No\\    
03 & G18.8 +0.3  & 12.1\tablenotemark{H}\tablenotemark{M} & 20   & 7.6  & 214 &   3& & 21.35   &  F  &  $13.8 \pm 0.4  $  & Yes\\       
04 & G20.0 -0.2  & 4.5\tablenotemark{M}  & 66  & 8.5  & 220 & 4  & &  66.40  &  F   & $11.2 \pm 0.3  $  & Yes\\   
05 & G21.5 -0.9  & $4.7 \pm 0.4$\tablenotemark{H}  & 68   & 8  & 220 & 5 & & 67.79   &  N  & $4.4 \pm 0.2  $ & No\\    
06 & G21.8 -0.6  & 5.5\tablenotemark{H}\tablenotemark{M}  & 86   & 8.0 & 220 & 6A &  & 93.35   &  N  &$ 5.6 \pm 0.2   $  & Yes \\ 
   &              & 5.2\tablenotemark{H}\tablenotemark{M}    &  85   & 8.0 & 220 & 6B & & & & \\
07 & G22.7 -0.2  & $4.4 \pm 0.4$\tablenotemark{M}  &  77  & 8.31 & 241 &  7  & &  76.63  &  N  &  $4.7 \pm 0.2 $    & Yes\\
08 & G23.3 -0.3  & $3.9 - 4.5$\tablenotemark{H}\tablenotemark{M}  & 66 - 80   & 7.6 & 214 & 8 &  & 78.51   &   N &  $4.8 \pm 0.2  $ & Yes\\  
09 & G23.6 +0.3  & 6.9\tablenotemark{M} \tablenotemark{S}  &  \nodata  & \nodata  & \nodata & 9 & & $ 99.95$  & N  &  $5.9 \pm 0.2$ & Possible\\ 
10 & G27.4 +0.0  & $7.5 - 9.8$ \tablenotemark{H}   &  $ V_{TP} - 84 $  & 8.5  & 220 & 10 &  & 99.95     &  N  &  $ 5.8 \pm 0.3$ & No\\   
11 & G29.7 -0.3  & $5.1 - 7.5$\tablenotemark{H}\tablenotemark{M} & 95  - 102 & 7.6   & 220  & 11A & & 95.00   &  N  &  $ 5.6 \pm 0.3 $ & Yes\\
   &              & 10.6\tablenotemark{M}    &  54   &  8.0 & 220 & 11B & & & & &\\
12 & G32.8 -0.1  & 4.8\tablenotemark{H}\tablenotemark{M}  &  $\sim81$  & 8.0  & 220 & 12 &  & 81.81   &  N  &$ 4.8 \pm 0.3 $  & Yes\\
13 & G33.6 +0.1  & 7.1\tablenotemark{M}\tablenotemark{S} & \nodata   & \nodata  & \nodata & 9  &  &  57.90  & N   & $ 3.5 \pm 0.3 $ &  No \\
14 & G34.7 -0.4  & $2.5 - 2.6$\tablenotemark{H}   &  42  &  8.5 & 220 &  14  &  & 50.48   &  N  & $3.0 \pm 0.3 $  & No\\
15 & G35.6 -0.4  & $3.6 \pm 0.4 $\tablenotemark{H}\tablenotemark{M}  & $\sim61$    & 8.5 &  220 & 15 & &  63.67  &   N & $ 3.8 \pm 0.3 $ & Possible\\
16 & G39.2 -0.3  & 6.2\tablenotemark{H}\tablenotemark{M} & 84    & 8.0  & 220 & 16 & & 69.39   & F   &  $8.5  \pm 0.5 $ & Yes\\
17 & G41.1 -0.3  & $ 8 - 9.7$\tablenotemark{H}  & \nodata   & 8.33  & 218 & 17 & &  $ V_{TP} - 63.01 $  & F  & $ 8.5 \pm 0.5$  & No\\
18 & G43.3 -0.2  & $8 - 11$\tablenotemark{H}  &  \nodata   &  \nodata  & \nodata & 18 & &  12.55  & F   & $11.3 \pm 0.4 $ & No\\ 
19 & G46.8 -0.3  & $6.8 - 8.6$\tablenotemark{H}  &  $ V_{TP} -  59$ & 10.0 & 250 & 19 & &  $V_{TP} - 0$ &  TP $-$ F & $5.7 \pm 0.9 - 11.4 \pm 0.5$ & No\\
20 & G49.2 -0.7  & 4.3\tablenotemark{H} & 70.7   & 8.4  & 254 & 20A & &    $V_{TP}$ & TP  & $5.4 \pm 0.6 $ & No\\
   &              & 6\tablenotemark{M} &  $ V_{TP}$   & 8.5  & \nodata &  20B   &        &     &     &        &    \\
21 & G54.1  +0.3  & $5.6 -7.2$\tablenotemark{H}\tablenotemark{M} &  $53\pm12$  & 7.6 & 220 & 21 & & 53.66  &  TP  & $4.9 \pm 0.8 $ & Yes\\
\enddata

\tablecomments{ 
\tablenotetext{a}{Literature distance estimation method - Superscript H: HI absorption, M: Molecular cloud association/interaction and S: $\Sigma$-D relation.}
\tablenotetext{b}{KDAR- Kinematic Distance Ambiguity Resolution, indicating whether the SNR is at the near (N), far (F) or tangent point (TP) distance.}
\tablenotetext{c} {Associated with $^{13}$CO.} 
\tablerefs { 
(1) \cite{Leahy2014}, (2) \cite{Johanson}, (3) \cite{Tian2007}, (4) \cite{Petriella2013}, (5) \cite{Tian2008}, (6A) \cite{Tian2008} , (6B) \cite{Zhou2009}, (7) \cite{Su2014}, (8) \cite{Leahy2008}, (9) \cite{Kilpatrick2016}, (10) \cite{TianLeahy2008}, (11A) \cite{LeahyTian2008}, (11B) \cite{Su2009}, (12) \cite{ZhouChen2011}, (14) \cite{Cox1999}, (15) \cite{Zhu2013}, (16) \cite{Su2011}, (17) \cite{LeahyRanasinghe2016}, (18) \cite{Brogan2001}, (19) \cite{Sato1979}, (20A) \cite{TianLeahy2013}, (20B) \cite{Koo1995}, (21) \cite{LeahyTianWang2008}.}}
\label{tab:table1}
\end{deluxetable*}

\section*{Acknowledgements}

This work was supported in part by a grant from the Natural Sciences and Engineering Research Council of Canada. 
% We would also like to thank the referee for the insightful comments and important suggestions that have improved this work.


\begin{thebibliography}{}

\bibitem[Anderson \& Bania(2009)]{Anderson22009} Anderson, L.~D., \& Bania, T.~M.\ 2009, \apj, 690, 706 

\bibitem[Anderson et al.(2009)]{Anderson20091} Anderson, L.~D., Bania, T.~M., Jackson, J.~M., et al.\ 2009, \apjs, 181, 255 

\bibitem[Becker \& Helfand(1985)]{BeckerHelfand1985} Becker, R.~H., \& Helfand, D.~J.\ 1985, \apjl, 297, L25

\bibitem[Brogan \& Troland(2001)]{Brogan2001} Brogan, C.~L., \& Troland, T.~H.\ 2001, \apj, 550, 799 

\bibitem[Caswell et al.(1975)]{Caswell1975} Caswell, J.~L., Murray, J.~D., Roger, R.~S., Cole, D.~J., \& Cooke, D.~J.\ 1975, \aap, 45, 239 

\bibitem[Cox(1972)]{1972Cox} Cox, D.~P.\ 1972, \apj, 178, 159

\bibitem[Cox et al.(1999)]{Cox1999} Cox, D.~P., Shelton, R.~L., Maciejewski, W., et al.\ 1999, \apj, 524, 179 

\bibitem[Cox(2005)]{2005Cox} Cox, D.~P.\ 2005, \araa, 43, 337 

\bibitem[Ferri{\`e}re(2001)]{2001Ferriere} Ferri{\`e}re, K.~M.\ 2001, Reviews of Modern Physics, 73, 1031

\bibitem[Frail \& Clifton(1989)]{Frail1989} Frail, D.~A., \& Clifton, T.~R.\ 1989, \apj, 336, 854

\bibitem[Giacani et al.(2009)]{Giacani2009} Giacani, E., Smith, M.~J.~S., Dubner, G., et al.\ 2009, \aap, 507, 841 

\bibitem[Gotthelf \& Vasisht(1997)]{Gotthelf1997} Gotthelf, E.~V., \& Vasisht, G.\ 1997, \apjl, 486, L133 

\bibitem[Green(1984)]{Green1984da} Green, D.~A.\ 1984, \mnras, 209, 449

\bibitem[Green(1989)]{Green1989DA} Green, D.~A.\ 1989, \mnras, 238, 737 

\bibitem[Green(2014)]{Green2014} Green, D.~A.\ 2014, Bulletin of the Astronomical Society of India, 42, 47

\bibitem[Green(2015)]{2015Green} Green, D.~A.\ 2015, \mnras, 454, 1517 

\bibitem[Green(2017)]{2017Green} Green, D.~A.\ 2017, VizieR Online Data Catalog, 7278

\bibitem[Gwinn et al.(1992)]{Gwinn1992} Gwinn, C.~R., Moran, J.~M., \& Reid, M.~J.\ 1992, \apj, 393, 149 

\bibitem[Hewitt et al.(2009)]{HewittRho2009} Hewitt, J.~W., Rho, J., Andersen, M., \& Reach, W.~T.\ 2009, \apj, 694, 1266 

\bibitem[Ilovaisky \& Lequeux(1972)]{Ilovaisky1972I} Ilovaisky, S.~A., \& Lequeux, J.\ 1972, \aap, 18, 169 

\bibitem[Jackson et al.(2006)]{Jackson} Jackson, J.~M., Rathborne, J.~M., Shah, R.~Y., et al.\ 2006, \apjs, 163, 145 

\bibitem[Johanson \& Kerton(2009)]{Johanson} Johanson, A.~K., \& Kerton, C.~R.\ 2009, \aj, 138, 1615 

\bibitem[Jones \& Dickey(2012)]{Jones2012} Jones, C., \& Dickey, J.~M.\ 2012, \apj, 753, 62 

\bibitem[Kilpatrick et al.(2016)]{Kilpatrick2016} Kilpatrick, C.~D., Bieging, J.~H., \& Rieke, G.~H.\ 2016, \apj, 816, 1

\bibitem[Koo \& Moon(1997)]{Koo1997} Koo, B.-C., \& Moon, D.-S.\ 1997, \apj, 475, 194 

\bibitem[Koo et al.(1995)]{Koo1995} Koo, B.-C., Kim, K.-T., \& Seward, F.~D.\ 1995, \apj, 447, 211 

\bibitem[Kwee et al.(1954)]{1954kwee} Kwee, K.~K., Muller, C.~A., \& Westerhout, G.\ 1954, \bain, 12, 211 

\bibitem[Leahy \& Tian(2008)]{Leahy2008} Leahy, D.~A., \& Tian, W.~W.\ 2008, \aj, 135, 167 

\bibitem[Leahy \& Tian(2008)]{LeahyTian2008} Leahy, D.~A., \& Tian, W.~W.\ 2008, \aap, 480, L25 

\bibitem[Leahy \& Tian(2010)]{Leahy2010} Leahy, D., \& Tian, W.\ 2010, The Dynamic Interstellar Medium: A Celebration of the Canadian Galactic Plane Survey, 438, 365 

\bibitem[Leahy et al.(2008)]{LeahyTianWang2008} Leahy, D.~A., Tian, W., \& Wang, Q.~D.\ 2008, \aj, 136, 1477 

\bibitem[Leahy et al.(2014)]{Leahy2014} Leahy, D., Green, K., \& Tian, W.\ 2014, \mnras, 438, 1813 

\bibitem[Leahy \& Ranasinghe(2016)]{LeahyRanasinghe2016} Leahy, D.~A., \& Ranasinghe, S.\ 2016, \apj, 817, 74

\bibitem[Leahy \& Williams(2017)]{2017LeahyWilliams} Leahy, D.~A., \& Williams, J.~E.\ 2017, \aj, 153, 239

\bibitem[Leahy(2017)]{2017Leahy} Leahy, D.~A.\ 2017, \apj, 837, 36 

\bibitem[Lockhart \& Goss(1978)]{LockhartGoss1978} Lockhart, I.~A., \& Goss, W.~M.\ 1978, \aap, 67, 355 

\bibitem[Lu et al.(2002)]{LU2002} Lu, F.~J., Wang, Q.~D., Aschenbach, B., Durouchoux, P., \& Song, L.~M.\ 2002, \apjl, 568, L49 

\bibitem[Moffett \& Reynolds(1994)]{Moffeett1994} Moffett, D.~A., \& Reynolds, S.~P.\ 1994, \apj, 437, 705

\bibitem[Patnaik et al.(1990)]{Patnaik1990} Patnaik, A.~R., Hunt, G.~C., Salter, C.~J., Shaver, P.~A., \& Velusamy, T.\ 1990, \aap, 232, 467 

\bibitem[Persic et al.(1996)]{Persic} Persic, M., Salucci, P., \& Stel, F.\ 1996, \mnras, 281, 27

\bibitem[Petriella et al.(2013)]{Petriella2013} Petriella, A., Paron, S.~A., \& Giacani, E.~B.\ 2013, \aap, 554, A73 

\bibitem[Pinheiro Gon{\c c}alves et al.(2011)]{Goncalves2011} Pinheiro Gon{\c c}alves, D., Noriega-Crespo, A., Paladini, R., Martin, P.~G., \& Carey, S.~J.\ 2011, \aj, 142, 47 

\bibitem[Quireza et al.(2006)]{Quizera2006B} Quireza, C., Rood, R.~T., Balser, D.~S., \& Bania, T.~M.\ 2006, \apjs, 165, 338 

\bibitem[Quireza et al.(2006)]{Quizera2006A} Quireza, C., Rood, R.~T., Bania, T.~M., Balser, D.~S., \& Maciel, W.~J.\ 2006, \apj, 653, 1226 

\bibitem[Radhakrishnan et al.(1972)]{Radha1972} Radhakrishnan, V., Goss, W.~M., Murray, J.~D., \& Brooks, J.~W.\ 1972, \apjs, 24, 49 

\bibitem[Ranasinghe \& Leahy(2017)]{2017Ranasinghe} Ranasinghe, S., \& Leahy, D.~A.\ 2017, \apj, 843, 119 

\bibitem[Ranasinghe \& Leahy(2017)]{2017RanasingheLeahy} Ranasinghe, S., \& Leahy, D.~A.\ 2017, arXiv:1712.04423 

\bibitem[Ranasinghe et al.(2017)]{2017RLT} Ranasinghe, S., Leahy, D.~A., \& Tian, W.\ 2017, arXiv:1712.04515

\bibitem[Reid et al.(2014)]{Reid2014} Reid, M.~J., Menten, K.~M., Brunthaler, A., et al.\ 2014, \apj, 783, 130 

\bibitem[Sanbonmatsu \& Helfand(1992)]{Sanbonmatsu1992} Sanbonmatsu, K.~Y., \& Helfand, D.~J.\ 1992, \aj, 104, 2189

\bibitem[Sato(1973)]{Sato1973} Sato, F.\ 1973, \pasj, 25, 135 

\bibitem[Sato(1979)]{Sato1979} Sato, F.\ 1979, \aplett, 20, 43 

\bibitem[Shaver \& Goss(1970)]{Shaver1970} Shaver, P.~A., \& Goss, W.~M.\ 1970, Australian Journal of Physics Astrophysical Supplement, 14, 133

\bibitem[Stil et al.(2006)]{Stil} Stil, J.~M., Taylor, A.~R., Dickey, J.~M., et al.\ 2006, \aj, 132, 1158 

\bibitem[Su et al.(2009)]{Su2009} Su, Y., Chen, Y., Yang, J., et al.\ 2009, \apj, 694, 376 

\bibitem[Su et al.(2011)]{Su2011} Su, Y., Chen, Y., Yang, J., et al.\ 2011, \apj, 727, 43 

\bibitem[Su et al.(2014)]{Su2014} Su, Y., Yang, J., Zhou, X., Zhou, P., \& Chen, Y.\ 2014, \apj, 796, 122 

\bibitem[Tian \& Leahy(2008)]{Tian2008} Tian, W.~W., \& Leahy, D.~A.\ 2008, \mnras, 391, L54

\bibitem[Tian \& Leahy(2008)]{TianLeahy2008} Tian, W.~W., \& Leahy, D.~A.\ 2008, \apj, 677, 292-296

\bibitem[Tian \& Leahy(2013)]{TianLeahy2013} Tian, W.~W., \& Leahy, D.~A.\ 2013, \apjl, 769, L17 

\bibitem[Tian et al.(2007)]{Tian2007} Tian, W.~W., Leahy, D.~A., \& Wang, Q.~D.\ 2007, \aap, 474, 541 

\bibitem[Truelove \& McKee(1999)]{1999TM} Truelove, J.~K., \& McKee, C.~F.\ 1999, \apjs, 120, 299 

\bibitem[Vasisht et al.(2000)]{Vasisht} Vasisht, G., Gotthelf, E.~V., Torii, K., \& Gaensler, B.~M.\ 2000, \apjl, 542, L49 

\bibitem[Willis(1973)]{Willis1973} Willis, A.~G.\ 1973, \aap, 26, 237 

\bibitem[Wood \& Churchwell(1989)]{Wood1989} Wood, D.~O.~S., \& Churchwell, E.\ 1989, \apjs, 69, 831 

\bibitem[Zhang et al.(2017)]{2017Zhang} Zhang, M.~F., Tian, W.~W., Leahy, D.~A., et al.\ 2017, \apj, 849, 147

\bibitem[Zhou \& Chen(2011)]{ZhouChen2011} Zhou, P., \& Chen, Y.\ 2011, \apj, 743, 4 

\bibitem[Zhou et al.(2009)]{Zhou2009} Zhou, X., Chen, Y., Su, Y., \& Yang, J.\ 2009, \apj, 691, 516 

\bibitem[Zhu et al.(2013)]{Zhu2013} Zhu, H., Tian, W.~W., Torres, D.~F., Pedaletti, G., \& Su, H.~Q.\ 2013, \apj, 775, 95 

\end{thebibliography}
\end{document}